\documentclass[twocolumn,showpacs,prb,superscriptaddress]{revtex4}

\usepackage{graphicx}
\usepackage{color}
\usepackage{tabularx}
\usepackage{epsfig}
\usepackage{amsmath}
\usepackage{amssymb}
\usepackage{graphicx}
\usepackage{dcolumn}
\usepackage{bm}
\usepackage{wasysym}

\definecolor{grn}{rgb}{0,0,0.54}

\newcommand{\bra}[1]{\langle #1|}
\newcommand{\ket}[1]{|#1\rangle}

\newcommand{\qdmlr}{\unitlength 0.03em
  \begin{minipage}{35\unitlength}
    \begin{center}
      \begin{picture}(30,17)
        \put(0,0){\linethickness{0.20mm}\line(1,0){20}}
	\put(10,17.3){\linethickness{0.20mm}\line(1,0){20}}
      \end{picture}
    \end{center}
  \end{minipage}
}

\newcommand{\qdmud}{\unitlength 0.03em 
  \begin{minipage}{35\unitlength}
    \begin{center}
      \begin{picture}(30,17)
        \linethickness{0.20mm}\qbezier(0,0)(5,8.65)(10,17.3)
        \linethickness{0.20mm}\qbezier(20,0)(25,8.65)(30,17.3)
      \end{picture}
    \end{center}
  \end{minipage}
}

\newcommand{\plaqueteA}{\unitlength 0.02em 
  \begin{minipage}{35\unitlength}
    \begin{center}
      \begin{picture}(30,17)
        \linethickness{0.14mm}\qbezier(0,0)(5,8.65)(10,17.3)
        \linethickness{0.14mm}\qbezier(20,0)(25,8.65)(30,17.3)
        \put(0,0){\linethickness{0.14mm}\line(1,0){20}}
        \put(10,17.3){\linethickness{0.14mm}\line(1,0){20}}
      \end{picture}
    \end{center}
  \end{minipage}
}

\newcommand{\plaqueteB}{\unitlength 0.02em 
  \begin{minipage}{35\unitlength}
    \begin{center}
      \begin{picture}(30,17)
        \linethickness{0.14mm}\qbezier(10,0)(5,8.65)(0,17.3)
        \linethickness{0.14mm}\qbezier(30,0)(25,8.65)(20,17.3)
        \put(10,0){\linethickness{0.14mm}\line(1,0){20}}
        \put(0,17.3){\linethickness{0.14mm}\line(1,0){20}}
      \end{picture}
    \end{center}
  \end{minipage}
}

\newcommand{\plaqueteC}{\unitlength 0.02em 
  \begin{minipage}{35\unitlength}
    \begin{center}
      \begin{picture}(30,35)
        \linethickness{0.14mm}\qbezier(10,0)(5,8.65)(0,17.3)
        \linethickness{0.14mm}\qbezier(20,17.3)(20,17.3)(10,34.6)
        \linethickness{0.14mm}\qbezier(0,17.3)(0,17.3)(10,34.6)
        \linethickness{0.14mm}\qbezier(10,0)(15,8.65)(20,17.3)
      \end{picture}
    \end{center}
  \end{minipage}
}

\begin{document}

\title{Engineering exotic phases for topologically-protected quantum 
computation by emulating quantum dimer models}

\author{A.~Fabricio Albuquerque}
\affiliation{Theoretische Physik, ETH Zurich, 8093 Zurich, Switzerland}
\affiliation{School of Physics, The University of New South Wales, Sydney,
New South Wales 2052, Australia}

\author{Helmut G.~Katzgraber}
\author{Matthias Troyer}
\author{Gianni Blatter}
\affiliation{Theoretische Physik, ETH Zurich, 8093 Zurich, Switzerland}

\date{\today}

\begin{abstract}

We use a nonperturbative extended contractor renormalization (ENCORE)
method for engineering quantum devices for the implementation of
topologically protected quantum bits described by an effective quantum
dimer model on the triangular lattice. By tuning the couplings of the
device, topological protection might be achieved if the ratio between
effective two-dimer interactions and flip amplitudes lies in the liquid
phase of the phase diagram of the quantum dimer model. For a proposal
based on a quantum Josephson junction array [L.~B.~Ioffe {\it et al.},
Nature (London) {\bf 415}, 503 (2002)] our results show that optimal
operational temperatures below 1 mK can only be obtained if extra
interactions and dimer flips, which are not present in the standard
quantum dimer model and involve three or four dimers, are included.
It is unclear if these extra terms in the quantum dimer Hamiltonian
destroy the liquid phase needed for quantum computation. Minimizing
the effects of multi-dimer terms would require energy scales in the
nano-Kelvin regime. An alternative implementation based on cold atomic
or molecular gases loaded into optical lattices is also discussed,
and it is shown that the small energy scales involved---implying
long operational times---make such a device impractical. Given the
many orders of magnitude between bare couplings in devices, and the
topological gap, the realization of topological phases in quantum
devices requires careful engineering and large bare interaction scales.

\end{abstract}

\maketitle

\section{Introduction}
\label{sec:introduction}

Systems characterized by topological quantum order (TQO)
have a degenerate ground state, which is not associated with
any broken symmetry, i.e., the different degenerate ground
states are indistinguishable under the action of any local
operator.\cite{alet:06} Instead, they can only be distinguished
via global operators intimately related to their topological
properties. TQO does not fit into Landau's paradigm for
ordered phases of matter,\cite{landau:37} which makes it intrinsically
interesting. Furthermore, this robustness against local perturbations
characteristic of systems exhibiting TQO can be used to implement a
fault-tolerant quantum computer.\cite{kitaev:03}

Within this approach, robust storage devices for quantum states
(``protected memory qubits'') can be built from Abelian topological
quantum states, whereas topologically-protected computations
(``protected gates'') can be implemented using non-Abelian
states.\cite{kitaev:03} Given the enormous challenges involved in
building conventional quantum computers caused by the decoherence
inherent to quantum-mechanical systems, the alternative approach
exploiting topological order has attracted considerable interest
recently because local operators (i.e., noise) do not disturb the
topological phase.

One promising class of systems exhibiting TQO are fractional
quantum Hall systems with filling factors $\nu = 5/2$ and
$\nu = 12/5$ which are conjectured to exhibit non-Abelian
anyonic excitations.\cite{moore:91} Unfortunately, despite
some evidence,\cite{camino:05} the existence of anyons in
these systems remains to be confirmed experimentally.  On the
other hand, a number of interesting lattice models is known
to exhibit TQO. Among these are quantum dimer models (QDM)
(Refs.~\onlinecite{rokhsar:88,moessner:01,ioffe:02,misguich:02})
spin models, and Hubbard models with generalized interactions defined
on Kagome lattices,\cite{balents:02,freedman:05,sheng:05,isakov:06}
toric,\cite{kitaev:03} and color\cite{bombin:06} codes, as well
as Kitaev's honeycomb anisotropic spin model.\cite{kitaev:06a} In
general, these lattice models incorporate unrealistic elements such as
artificially-constrained degrees of freedom or nontrivial interactions
and thus experimental realizations remain elusive.  Therefore,
we are interested in engineering topologically-ordered phases by
{\em emulating} lattice models using highly manipulable quantum {\em
tool-boxes}, such as Josephson junction arrays\cite{ioffe:02} and cold
atomic\cite{buechler:05} or molecular\cite{micheli:06,pupillo:08} gases
loaded into optical lattices. However, as promising as these approaches
might seem, the challenges imposed to the engineering of such emulators
are huge, requiring special attention to the {\it design} of such
devices and a careful analysis of the involved energy scales as well
as the possible existence of extra terms in the emulated Hamiltonian.

Having these issues in mind, we use a {\em nonperturbative}
algorithm, extended contractor renormalization
(ENCORE) (Ref.~\onlinecite{albuquerque:08b}) an
extended version of the Contractor Renormalization (CORE)
technique\cite{morningstar:94,morningstar:96} to design exotic
phases to build topological quantum computers as well as to propose
controllable experiments to investigate TQO.  We consider an
emulator for the QDM on the triangular lattice based on an array
of quantum Josephson junctions.\cite{ioffe:02} This system is a
good candidate for the implementation of a topologically-protected
qubit for two reasons: First, quantum dimer models are among
the best understood systems exhibiting TQO and the presence of a
topological phase has been unequivocally established in a number of
studies.\cite{moessner:01,ioffe:02,ralko:05,vernay:06,ralko:06} Second,
the manipulation of Josephson junctions is an experimentally mature
field where an exquisite degree of control has been achieved. We
are able to derive the couplings in the effective model describing
the low-energy physics in the array in an unbiased way (the only
limitations being caused by the finite sizes of the clusters
analyzed). Our final conclusion is that although the approach of
Ref.~\onlinecite{ioffe:02} seems promising based on simple estimates,
the energy scales obtained in the full analysis are too low to make
this approach feasible.

In addition, we also discuss, by means of a perturbative analysis,
an implementation based on cold atomic/molecular gases loaded into a
Kagome-shaped optical lattice and encounter similar problems of too
low energy scales and too long time scales.

\section{Devices for Emulating Quantum Dimer Models}
\label{sec:engineering}

\subsection{Quantum Dimer Model on a Triangular Lattice}
\label{subsec:qdm}

The QDM has first been introduced by Rokhsar and
Kivelson\cite{rokhsar:88} in the context of the resonating valence
bond (RVB) scenario for cuprate superconductors.\cite{anderson:87}
The square lattice version of this model only displays valence
bond crystal phases, with the notable exception of a single point
at which the correlations decay algebraically with distance and the
ground state splits into many topological sectors.\cite{rokhsar:88}
Its triangular lattice version, first analyzed by Moessner and
Sondhi,\cite{moessner:01} has a gapped liquid phase with exponentially
decaying correlations extending through a finite range of the model
parameters.

The triangular-lattice QDM is given by the following Hamiltonian
\begin{equation}
{\mathcal H} = 	{\mathcal H}_{\plaqueteA} + 
		{\mathcal H}_{\plaqueteB} + 
		{\mathcal H}_{\plaqueteC} ,
\label{eq:qdm}
\end{equation}
with
\begin{equation}
\begin{split}
{{\mathcal H}_{\plaqueteA}} = 
    -t \sum_{\plaqueteA} \big[
	\ket{\qdmlr} \bra{\qdmud} + 
	\ket{\qdmud} \bra{\qdmlr}
    \big] \\
    +v \sum_{\plaqueteA} \big[
	\ket{\qdmlr} \bra{\qdmlr}  + 
	\ket{\qdmud} \bra{\qdmud}
    \big]
\end{split}
\label{eq:qdm_plaq}
\end{equation}
and similar definitions for ${\mathcal H}_{\plaqueteB}$ and ${\mathcal
H}_{\plaqueteC}$.  Parallel dimers sitting on the same rhombus
(henceforth we refer to such configurations as {\em flippable rhombi})
flip with amplitude $t$ and interact with each other via a potential
strength $v$; the sum runs over all the rhombi with a given orientation.

Despite its apparent simplicity, the phase diagram of the QDM on
the triangular lattice is rich, comprising different crystalline
phases.\cite{moessner:01,ioffe:02,ralko:05,vernay:06,ralko:06}
Here we are only interested in the quantum liquid phase, which is
stabilized in the range $0.82 \lesssim v/t \le 1$,\cite{ralko:06}
with exponentially-decaying correlations between dimers and
a gap  $\Delta \sim 0.1t$ against excitations.\cite{ioffe:02}
In this phase the system's ground state is degenerate: twofold
degeneracy on a cylindrical geometry, fourfold on a torus (full
periodic boundary conditions). The topological sector to which a
given dimer configuration belongs can be determined via the parity
of the dimer count along an arbitrarily chosen reference line (see
Fig.~\ref{fig:jjk}), a property which can be used to build a two-level
system for a topologically protected quantum bit.\cite{ioffe:02}
Note that the topologically-ordered phase of the QDM is also stable
towards the presence of disorder,\cite{ioffe:02} a particularly useful
feature since the presence of imperfections would be unavoidable in
any putative engineered device.

\subsection{Emulator Based on Josephson Junctions}
\label{subsec:jja}

The emulation of the quantum dimer model on the triangular
lattice can be achieved by using Josephson junction arrays.
Ioffe {\em et al.}\cite{ioffe:02} introduced two different Josephson
junction array emulators for the QDM. In this work we discuss the
implementation defined on the Kagome lattice only, since it has
a smaller number of superconducting islands attached to each site
of the underlying triangular lattice and thus is more amenable to
numerical studies. However, our main conclusions are immediately
extended to the alternative implementation on a decorated triangular
lattice.

\begin{figure}[!tbp]
\includegraphics[width=0.35\textwidth,angle=270]{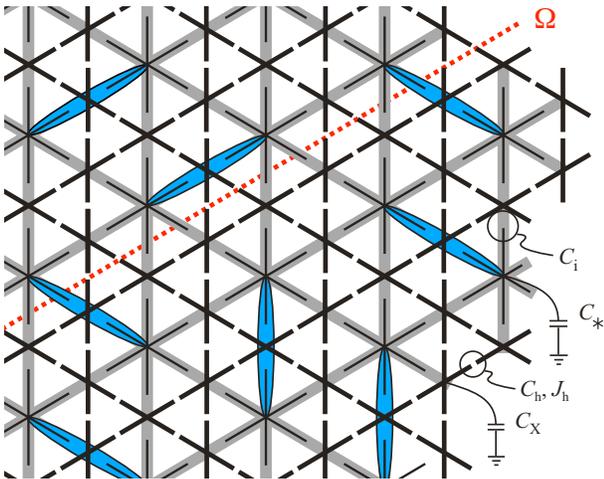}

\caption{
(Color online)
Array of Josephson junctions used to emulate the quantum dimer model
on the triangular lattice. The array
is formed by X-shaped superconducting islands (thick black lines),
which form a Kagome lattice and normal-state star-shaped islands
(thin black lines) placed at the center of every hexagon of the
Kagome lattice. The shaded lines are guides to the eye to emphasize
the underlying triangular lattice of the effective QDM. Cooper pairs
hop between nearest-neighbors X-shaped islands with an amplitude
given by the Josephson current $J_{\rm h}$. A large ratio between the
capacitances $C_{\rm i}$ and $C_{\rm h}$ defines a sizable on-hexagon
repulsion $E_{\rm hex}$ to emulate the hard-core dimer constraint.
The dimers are represented by ellipses sitting on one of the six
links of a given star-shaped island. The parity of the dimer count
along a reference line $\Omega$ (dotted line) is invariant under the
dimer flips in the Hamiltonian [Eq.~(\ref{eq:qdm})].
}
\label{fig:jjk}
\end{figure}

The proposed emulator is built from an array of X-shaped
superconducting islands structured as a kagome lattice, see
Fig.~\ref{fig:jjk} (thick black lines). Each X-shaped island is
coupled to its four neighboring islands by a capacitance $C_{\rm h}$
and a Josephson current $J_{\rm h}$. Inside every hexagon of the kagome
lattice a star-shaped island (thin black lines in Fig.~\ref{fig:jjk})
is placed which couples only capacitively to the X-shaped islands via
the capacitance $C_{\rm i}$.  The ground capacitance of an X-shaped
island is $C_{\rm X}$, whereas the ground capacitance of a star-shaped
island is $C_{\ast}$.  The energies associated with these couplings
are given by
\begin{equation}
E^{C} = \frac{(2e)^{2}}{2C}\,
\label{eq:coul_energy}
\end{equation}
where $e$ denotes the elementary charge, and $E^{J} = \hbar J/2e$.
We set $\hbar = 2e = 1$.

One dimer in this array is equivalent to a Cooper pair sitting on
one of the six X-shaped islands surrounding a given star-shaped
island, each one corresponding to one of the links of the underlying
triangular lattice (see Fig.~\ref{fig:jjk}, shaded thick lines).
By applying a global bias to the array, only half of a Cooper pair is
made available per star-shaped island and, in order to impose the
dimer hard-core condition and emulate the QDM Hamiltonian, we need to
tune the different capacitances and Josephson currents.  In order to
guarantee that no hexagon can participate in the formation of more
than one dimer (represented by ellipses in Fig.~\ref{fig:jjk}), we
assign a large value to the capacitance $C_{\rm i}$ such that there
is a strong repulsion between Cooper pairs placed around the same
star-shaped island.

The energetic cost for placing two bosons around the same hexagon
$E_{\rm hex}$ defines the basic energy scale of the array.  It should
not be confused with the on-site repulsion between two Cooper pairs
sitting on the same X-shaped island. The parity of the dimer count
along the reference line $\Omega$ (dotted line in Fig.~\ref{fig:jjk})
is invariant under dimer flips (local perturbations) in the Hamiltonian
[see Eq.~(\ref{eq:qdm})] and allows for the determination of the
topological sectors necessary to define a qubit state.

\subsection{Emulator Based on Cold Atomic/Molecular Gases}
\label{subsec:coldgas}

We also consider an implementation of the QDM based on cold
atomic/molecular gases loaded into a Kagome optical lattice, which can
in principle be created by using three laser beams,\cite{damski:05}
with the following Hamiltonian:
\begin{equation}
\begin{split}
{\mathcal H} = 	\frac{U}{2} \sum_{i} n_{i}(1-n_i) + 
		\frac{E_{\hexagon}}{2} \sum_{\hexagon} 
		n_{\hexagon}(n_{\hexagon}-1) \\
		- J \sum_{\left\langle i,j \right\rangle}
		(b^{\dagger}_{i}b_{j} + b^{\dagger}_{j}b_{i}) .
\end{split}
\label{eq:cold}
\end{equation}
Here $n_{i} = b^{\dagger}_{i}b_{i}$ is the bosonic number operator
at the site $i$ of the Kagome lattice, $U$ is a repulsion between
two bosons sitting on the same site and $J$ is the hopping amplitude
between nearest-neighbors sites $\left\langle i,j \right\rangle$ in the
Kagome lattice. $E_{\hexagon}$ is the energy required for placing two
bosons on {\em different} sites around the same hexagon in the Kagome
lattice and enforces the hard-core dimer condition. $n_{\hexagon}$
is the number of bosons sitting around a given hexagon.  Due to the
short-ranged interactions between cold atomic gases, the engineering
of interaction terms as in Eq.~(\ref{eq:cold}) would likely be a
highly nontrivial task. One possible solution to this problem is
to use polar molecules\cite{micheli:06,pupillo:08} whose permanent
dipole moment permits long-range interactions.

\section{Effective Hamiltonians from the ENCORE Method}
\label{sec:core}

The CORE method was originally introduced by Morningstar and
Weinstein\cite{morningstar:94,morningstar:96} and since then has
been successfully applied to different problems in strongly correlated
systems.\cite{piekarewicz:97,altman:02,capponi:02,berg:03,capponi:04,budnik:04,
abendschein:07} For our application we use an extended version,
ENCORE,  suitable for constrained models, such as the quantum dimer
model.\cite{albuquerque:08b}

The fundamental idea behind CORE and ENCORE is to derive an effective
model describing the low-energy physics of a lattice Hamiltonian by
reducing the number of degrees of freedom.  The usefulness of the
method relies on a fast decay of the effective interactions for the
specific effective model, something which needs to be verified for
each case.  A large amount of physical intuition is required to obtain
physically sound results, which is one reason why CORE has not found
a more widespread use to date.

The effective Hamiltonian obtained with ENCORE generally includes
arbitrarily-ranged terms. Large couplings associated with long-range
terms indicate that the restricted subspace does not accurately
describe the low-energy behavior of the original model.  However, if we
are interested in engineering an emulation of a certain Hamiltonian,
the aforementioned problems are irrelevant because in this case the
effective model and the restricted Hilbert space are known {\it a
priori}. If the ENCORE method fails we simply conclude that emulation
is not possible.

The breakdown of the mapping is also signaled by the appearance of
``intruder'' states in the low-lying spectrum. These are states
with negligible overlap with  any of the desired low-energy states.
Since both aforementioned effects are correlated,\cite{albuquerque:08b}
we avoid the adoption of an arbitrarily-defined threshold value for
the long-range interactions and we define the breakdown of the mapping
onto a QDM as the point where a first intruder state appears in the
device's low-energy spectrum.

Since our primary goal in the present paper is to verify the
feasibility of a fault-tolerant quantum bit engineered from a system
with a topologically-ordered phase, the device's parameters must
be tuned in order to ensure that the emulated model has couplings
known to correspond to a quantum dimer liquid phase.  In addition,
a careful analysis of the involved energy scales is necessary in
order to avoid technological limitations.

\section{Emulating Quantum Dimer Models Using Josephson Junction Arrays}
\label{sec:results}

The array of Josephson junctions discussed in Sec.~\ref{subsec:jja} can 
be described by the following generalized Bose-Hubbard Hamiltonian
\begin{equation}
{\mathcal H} = \frac{1}{2}\sum_{j,k}n_{j}\hat{C}^{-1}_{j,k}n_{k} - 
J_{h} \sum_{\left\langle j,k \right\rangle}(b^{\dagger}_{j}b_{k} 
+ b^{\dagger}_{k}b_{j}) .
\label{eq:bose_hubb}
\end{equation}
The positions of the X-shaped islands in the array are denoted by the
indices $j$ and $k$. ${\left\langle j,k \right\rangle}$ represent
nearest neighbor (NN) sites in the Kagome lattice.  $n_{j}
= b^{\dagger}_{j}b_{j}$ is the bosonic occupation number at site
$\vec{r_j}$, $J_{\rm h}$ is the Josephson current between two X-shaped
islands.

$\hat{C}^{-1}$ is obtained by numerically inverting the capacitance
matrix $\hat{C}$ of the array.  The matrix elements connecting two
X-shaped islands in this matrix are given by
\begin{equation}
\hat{C}_{j,k} = [C_{\rm X} + \mu_j C_{\rm i} +  \nu_j C_{\rm h}] 
	  \delta_{\vec{r_j}, \vec{r_k}}
	+ C_{\rm h} \delta_{\vec{r_j}, \vec{r_k} + \hat{r}} , 
\label{eq:C1}
\end{equation}
where $\hat{r}$ connects NN sites in the Kagome lattice, $\mu$
is the number of hexagons a given X-shaped island joins [$\mu = 2$
for full periodic boundary conditions (PBC)] and $\nu$ is its number
of NN [$\nu = 4$ for PBC]. 

The normal-state star-shaped islands are only capacitively
connected to the X-shaped islands and their only role is to set
up the interactions in the Hamiltonian. The inverse $\hat{C}^{-1}$
appearing in the Hamiltonian is sensitive also to these interactions,
specified in the following.  Star-shaped islands sitting on the sites
$\vec{R_{\alpha}}$ and $\vec{R_{\beta}}$ of the underlying triangular
lattice contribute with,
\begin{equation}
\hat{C}_{\alpha,\beta} = 
	[C_{\ast} + 6C_{\rm i}] \delta_{\vec{R_{\alpha}} , \vec{R_{\beta}}}
\label{eq:C2}
\end{equation}
and the elements connecting X- and star-shaped islands are
\begin{equation}
\hat{C}_{j,\alpha} = C_{\rm i} \delta_{\vec{r_j} , \vec{r_{\alpha}} + \vec{s}} ,
\label{eq:C3}
\end{equation}
where $\vec{s}$ are the vectors connecting a star-shaped island
to the X-shaped islands surrounding it.

The energy $E_{\rm hex}$ to place two dimers on a hexagon can be
obtained from certain matrix elements of $\hat{C}^{-1}$. Quantum
fluctuations due to the Josephson coupling  $J_{\rm h}$ reduce this
bare value and we thus include them in second order in perturbation
theory in our discussions below. To ensure that we are allowed to
restrict the calculations to hard-core bosons, we have verified that
the on-site repulsion is larger than $J_{\rm h}$ by a factor of at
least $50$ for all sets of couplings in the array.

\begin{figure}[!tbp]
\includegraphics[width=0.35\textwidth,angle=270]{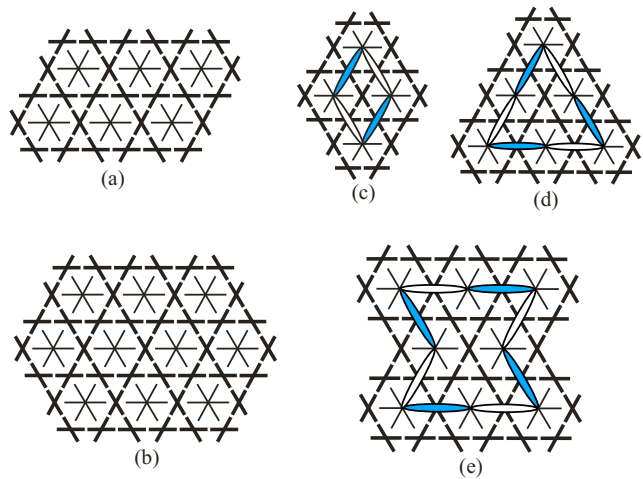}

\caption{(Color online) 
Open-boundary clusters studied: (a) $N\times2$ ($N=3$ in the figure)
hexagon ladders; (b) ten-hexagon cluster; (c) -- (e) special clusters
with four, six and eight hexagons,  accommodating the lowest-order flip
(represented by the associated transition graphs where full dimers
flip to open ones) involving two, three, and four dimers, respectively.
}
\label{fig:clusters}
\end{figure}

Our results are obtained by analyzing the small open-boundary
clusters depicted in Fig.~\ref{fig:clusters}: ladder-like clusters
with $N\times2$ hexagons ($N=3$, $4$, and $5$), a ten-hexagon cluster
(from which most results have been obtained), and three special clusters
with four, six, and eight hexagons which accommodate only two distinct
dimer configurations each.

Finally we need to take into account experimental limitations.
The smallest values for capacitances between two superconducting
islands obtainable with current technologies are such that $E^{C} =
1/2C \sim 1{\rm K}$ [see Eq.~(\ref{eq:coul_energy})], higher values of
$E^{C}$ can be obtained for ground capacitances.  We set the smallest
junction capacitance $C_{\rm h} = 0.5$, such that $E^{C}_{\rm h} = 1$
to set the energy scale, and we restrict our analysis to values of
$C_{\rm i} > C_{\rm h}$ throughout the rest of this paper.  In this
way, {\em assuming} a value of $E^{C}_{\rm h} \approx  1{\rm K}$,
{\em a priori} taking into account current technological limits,
all energies are fortuitously directly given in Kelvin.

\subsection{Dimer Flips}
\label{sec:flips}

The simplest dimer flip involves two parallel dimers on
the same rhombus of the triangular lattice, as illustrated in
Fig.~\ref{fig:clusters}(c). It involves the creation of a virtual state
in which one hexagon is doubly occupied, occurring with an amplitude
given in second-order perturbation theory by $t \approx J_{\rm
h}^{2} / E_{\rm hex}$.  Using ENCORE, we now analyze the amplitude
associated with this dimer move for the set of capacitances studied
in Ref.~\onlinecite{ioffe:02}:
\begin{eqnarray} 
C_{\ast}      = 10, && C_{\rm X}     = 10\,, \\ \nonumber 
C_{\rm i}  = 2.0, &&C_{\rm h}     = 0.5\,.  
\label{eq:ioffenums}
\end{eqnarray} 

\begin{figure}[!tbp]
\includegraphics*[width=0.35\textwidth,angle=270]{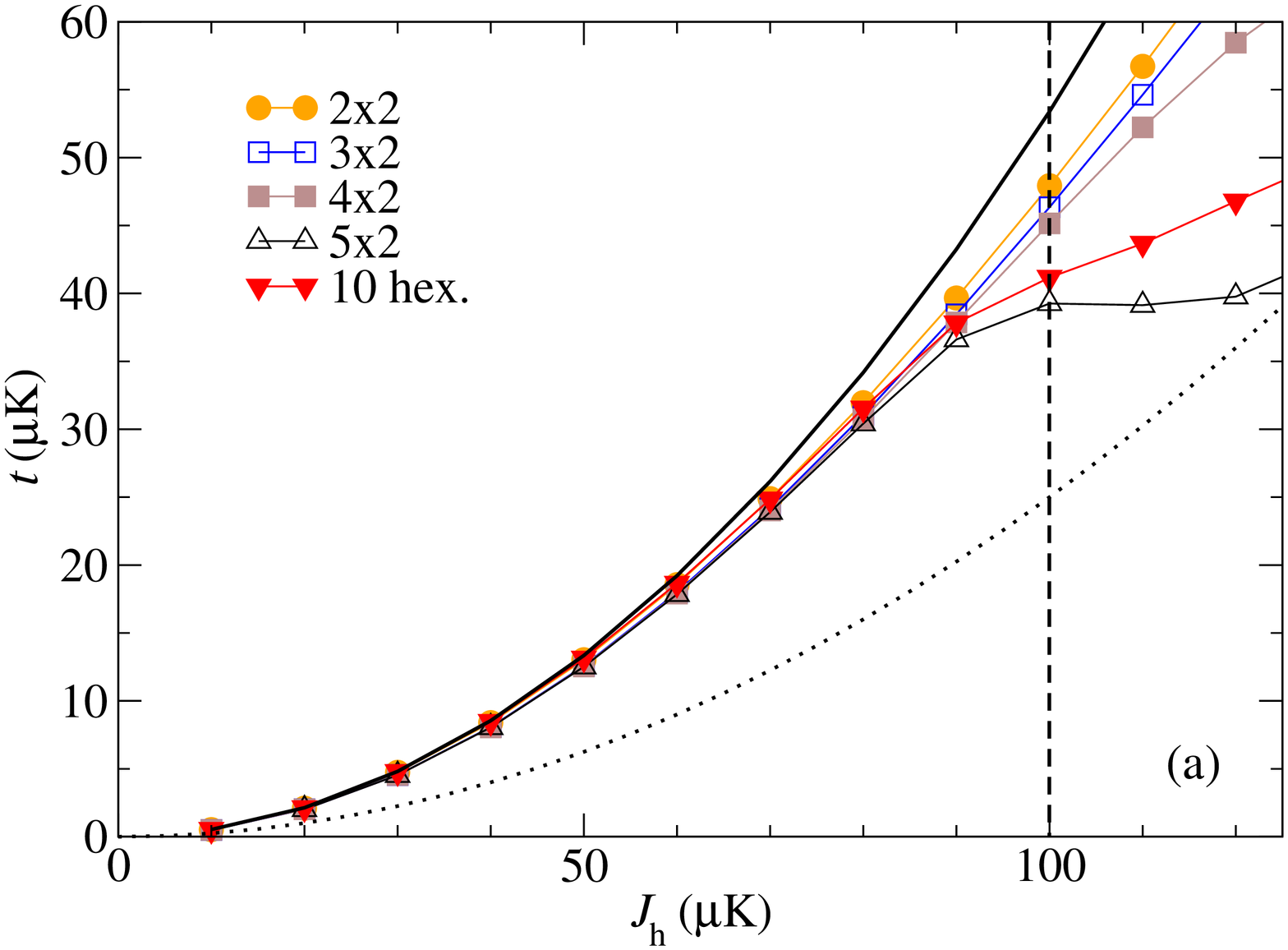}

\vspace{0.2cm}

\includegraphics*[width=0.35\textwidth,angle=270]{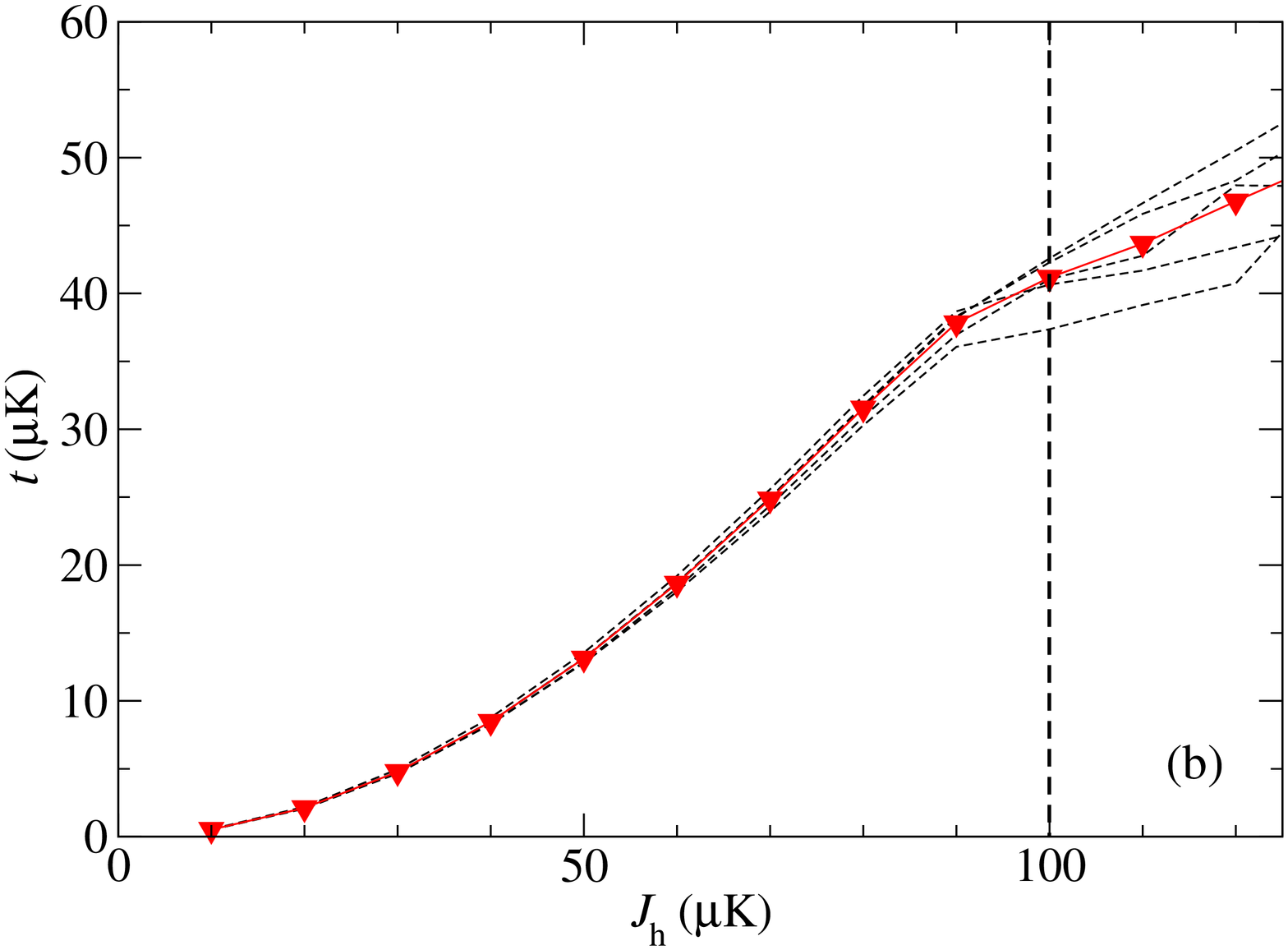}

\caption{(Color online) 
(a) Two-dimer flip amplitude $t$ calculated with ENCORE versus
the Josephson current $J_{\rm h}$ for the clusters depicted in
Fig.~\ref{fig:clusters}.  The parameters of Ref.~\onlinecite{ioffe:02}
are used: $C_{\ast} = C_{\rm X} = 10$, $C_{\rm i} = 2$, and $C_{\rm
h} = 0.5$. Solid lines are only guides to the eye. The vertical
dashed line indicates the point where the mapping onto a QDM breaks
down (cf. Sec.~\ref{sec:core}).  Second-order perturbative results
obtained by numerically calculating $E_{\rm hex}$ are indicated by
the thick black curve. The dotted curve corresponds to the results
obtained by using the expression for $E_{\rm hex}$ derived in
Ref.~\onlinecite{ioffe:02}.  (b) Amplitudes for each of the five
nonequivalent two-dimer flips in the ten-hexagon cluster (dashed
curves, see main text), compared to their average (downward triangles).
}
\label{fig:t_01}
\end{figure}

In Fig.~\ref{fig:t_01}(a) we show the results for the clusters depicted
in Figs.~\ref{fig:clusters}(a)--\ref{fig:clusters}(c).  The absence
of substantial finite-size effects confirms that the two-dimer flip
is a local process. The different clusters differ only after the point
 where the mapping onto a QDM breaks down due to the appearance
of intruder states (see Sec.~\ref{sec:core}).  The results from
the ten-hexagon cluster are obtained from an average between the
amplitudes for five possible nonequivalent two-dimer flips occurring
within slightly different ``dimer environments,'' the different
ways the dimers not participating in the flip are arranged in the
cluster.\cite{albuquerque:08b} Again, the amplitudes for these
individual processes only deviate slightly from their average
until the point where the dimer picture breaks down, as shown in
Fig.~\ref{fig:t_01}(b).

\begin{figure}[!tbp]
\includegraphics[width=0.35\textwidth,angle=270]{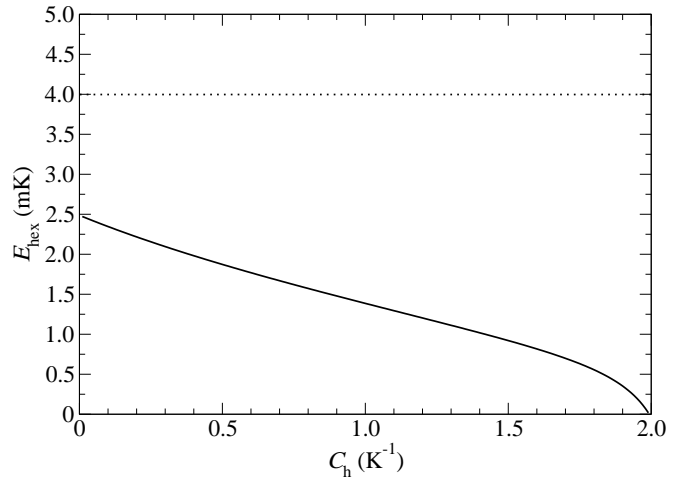}

\caption{ 
Numerical second-order perturbative results for the on-hexagon
repulsion $E_{\rm hex}$ for the ten-hexagon cluster as a function
of the capacitance $C_{\rm h}$ for $C_{\ast} = C_{\rm X} = 10$
and $C_{\rm i} = 2$. The horizontal dotted line is $E_{\rm hex}
\sim 0.2 (C_{\rm i} / C_{\rm X})^2 E^{C}_{\ast}$, as obtained in
Ref.~\onlinecite{ioffe:02}.
}
\label{fig:onhex}
\end{figure}

In Fig.~\ref{fig:t_01}(a) we also compare to second-order perturbative
data for $t$, obtained from $t =J_{\rm h}^{2} / E_{\rm hex}$ by
numerically calculating $E_{\rm hex}$ for the ten-hexagon cluster
(solid line; see Fig.~\ref{fig:onhex}), as well as by using the
approximation $E_{\rm hex} \sim 0.2 (C_{\rm i} / C_{\rm X})^2
E^{C}_{\ast}$ of Ref.~\onlinecite{ioffe:02}.  The discrepancy between
the two estimates clearly illustrates the nontrivial dependence
of $E_{\rm hex}$ on the set of capacitances adopted for the array.
Note that $E_{\rm hex}$ vanishes when $C_{\rm h}=C_{\rm i}$.  Using the
accurate estimate for $E_{\rm hex}$ (solid line in Fig.~\ref{fig:onhex}), we
find reasonable agreement  with the ENCORE results, which motivates
us to use the perturbative results to guide our optimizations below.

Additional flips involving three and four dimers occurring
within third and fourth order in the small parameter $J_{\rm h}$
with amplitudes $t_3$ and $t_4$ respectively, are depicted in
Figs.~\ref{fig:clusters}(d) and \ref{fig:clusters}(e). Together with
the two-dimer flip, these processes are special because they are the
lowest-order possible dimer moves in the array: comprising $m$ dimers,
they appear as $m$th order processes in $J_{\rm h}$. All other flip
terms are strongly suppressed in the limit of small currents $J_{\rm
h}$. Throughout this paper, we denote the sum of the {\em absolute
values} of the amplitudes for all other dimer moves by $\Sigma$, and
we use this quantity to gauge the validity of the mapping onto a QDM.

Of particular interest is the flip involving three dimers in a
triangular configuration depicted in Fig.~\ref{fig:clusters}(d), which
has also been found in a recent mean-field mapping by Vernay {\it et
al.}\cite{vernay:06}~of a spin-orbital model for the compound LiNiO$_2$
onto a QDM. They found that this extra dimer move considerably to
extend the liquid phase, allowing for ``extra room'' in trying to
optimize the couplings in the array.  One is tempted to conclude
that the four-dimer term $t_4$ has similar effects, and it will be
of interest to confirm this numerically.

\begin{figure}
\centering
\includegraphics[width=0.35\textwidth,angle=270]{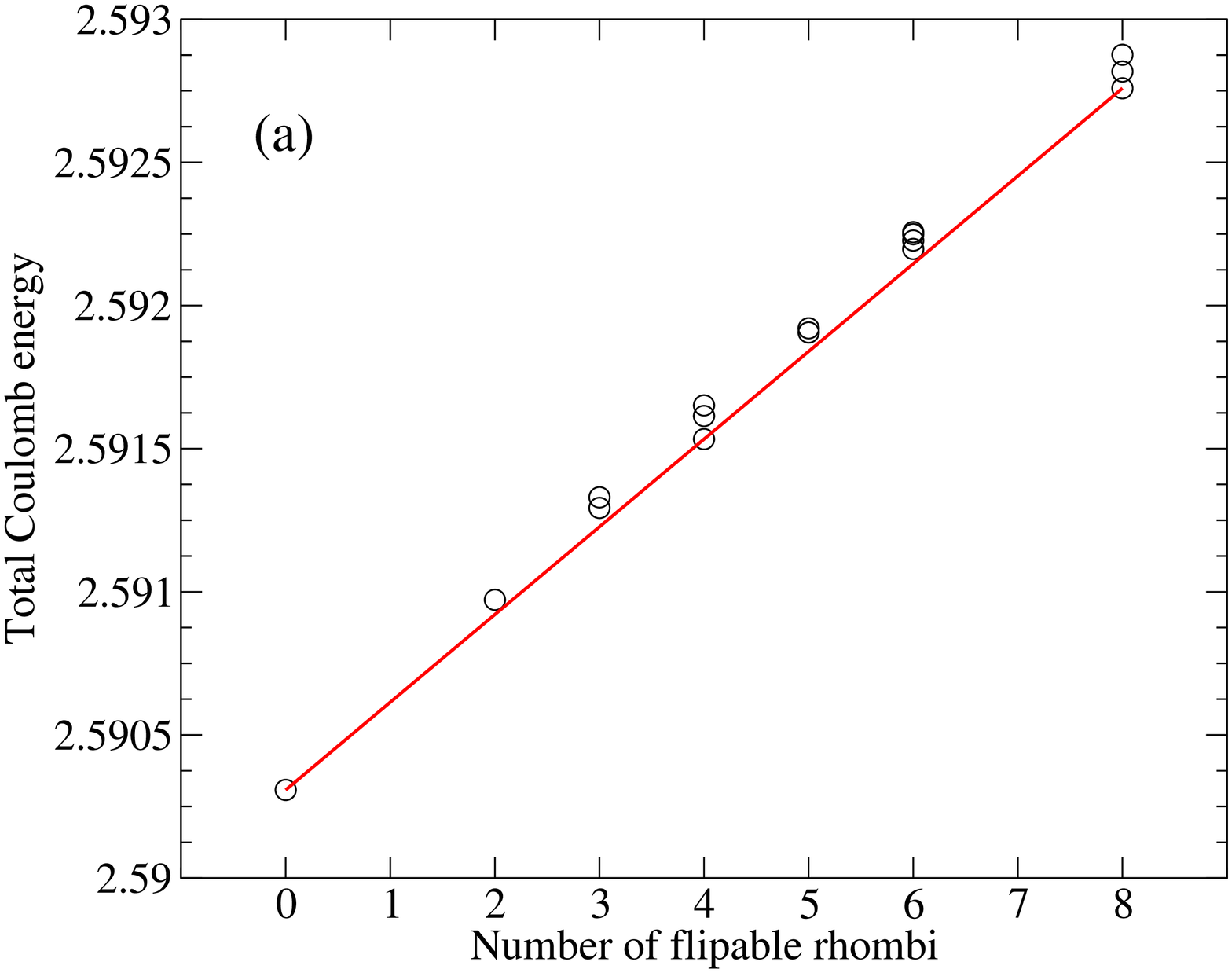}

\vspace{0.5cm}

\includegraphics[width=0.35\textwidth]{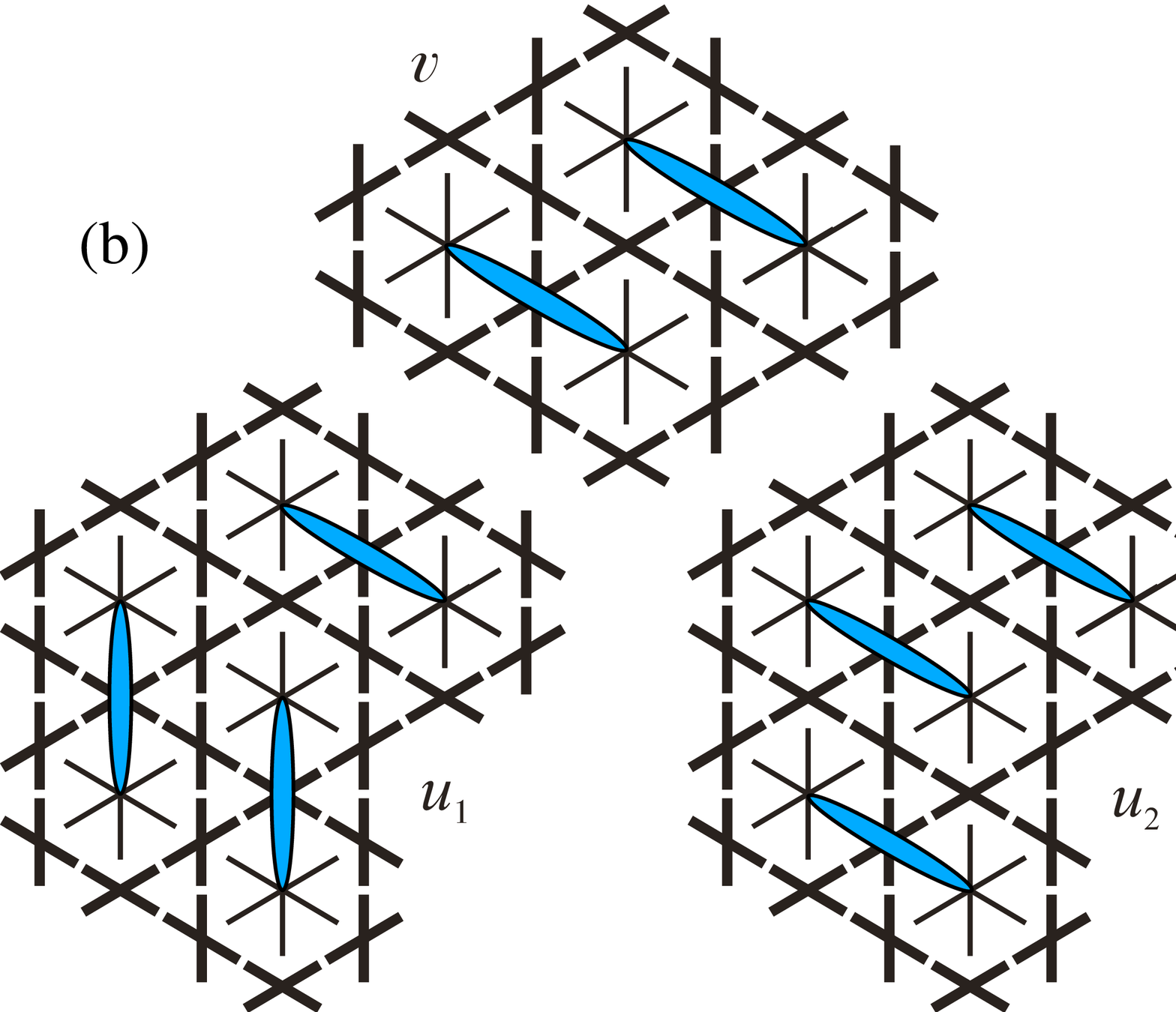}
\caption{(Color online) 
(a) Total Coulomb energy versus number of flippable rhombi (parallel dimers
which can flip) for a $4 \times 4$ PBC cluster in
the limit $J_{\rm h}=0$ and with $C_{\ast} = C_{\rm X} = 10$, $C_{\rm i} = 2$,
and $C_{\rm h} = 0.5$.  Deviations  from the dominant linear
behavior signal the presence of extra interaction terms beyond the
one between parallel dimers sitting on the same rhombus with strength
$v$. (b) Interactions present in the limit $J_{\rm h}=0$.
In addition to the dominant rhombus term $v$, two
further interaction terms involving three dimers with amplitudes
$u_1$ and $u_2$ have to be included. The deviations from the main
contribution in panel (a) can be explained by the presence of these
extra terms.
}
\label{fig:v_01}
\end{figure}

\subsection{Dimer Interactions}
\label{sec:interactions}

Dimer-dimer interactions have a nontrivial dependence on the particular
choice of capacitances in the array and must be tuned in order for
the ratio $v/t$ to lie in the liquid phase.  To investigate them, we
first analyze a $4 \times 4$ cluster with PBC in the limit of zero
Josephson current ($J_{\rm h} = 0$) by inverting the capacitance
matrix on this cluster.  In Fig.~\ref{fig:v_01}(a) we show the
Coulomb energy as a function of the number of flippable rhombi in
each configuration. The total energy scales well with the number of
flippable rhombi, confirming that the interaction between parallel
dimers sitting on the same rhombus $v$ is the dominant diagonal term
in the emulated dimer Hamiltonian.  The deviations from the linear
behavior in Fig.~\ref{fig:v_01}(a) indicate that other interaction
terms are also present. These extra contributions cannot be explained
by pairwise interactions, but {\it all} such deviations are entirely
described if we take into account three-dimer interactions with
strengths $u_1$ and $u_2$ [as shown in Fig.~\ref{fig:v_01}(b)].

\begin{figure*}[!tbp]
\includegraphics*[width=\columnwidth]{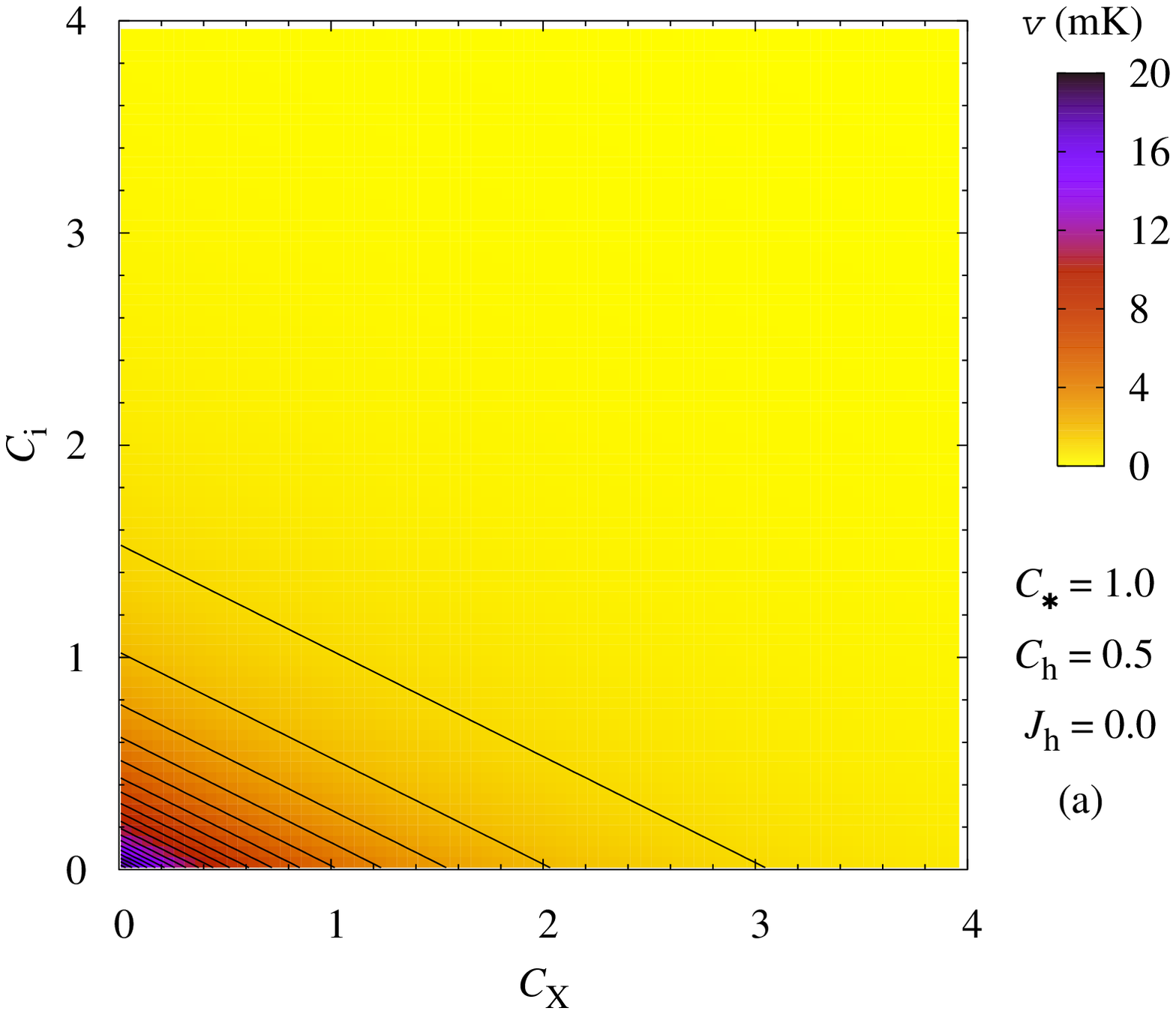}
\includegraphics*[width=\columnwidth]{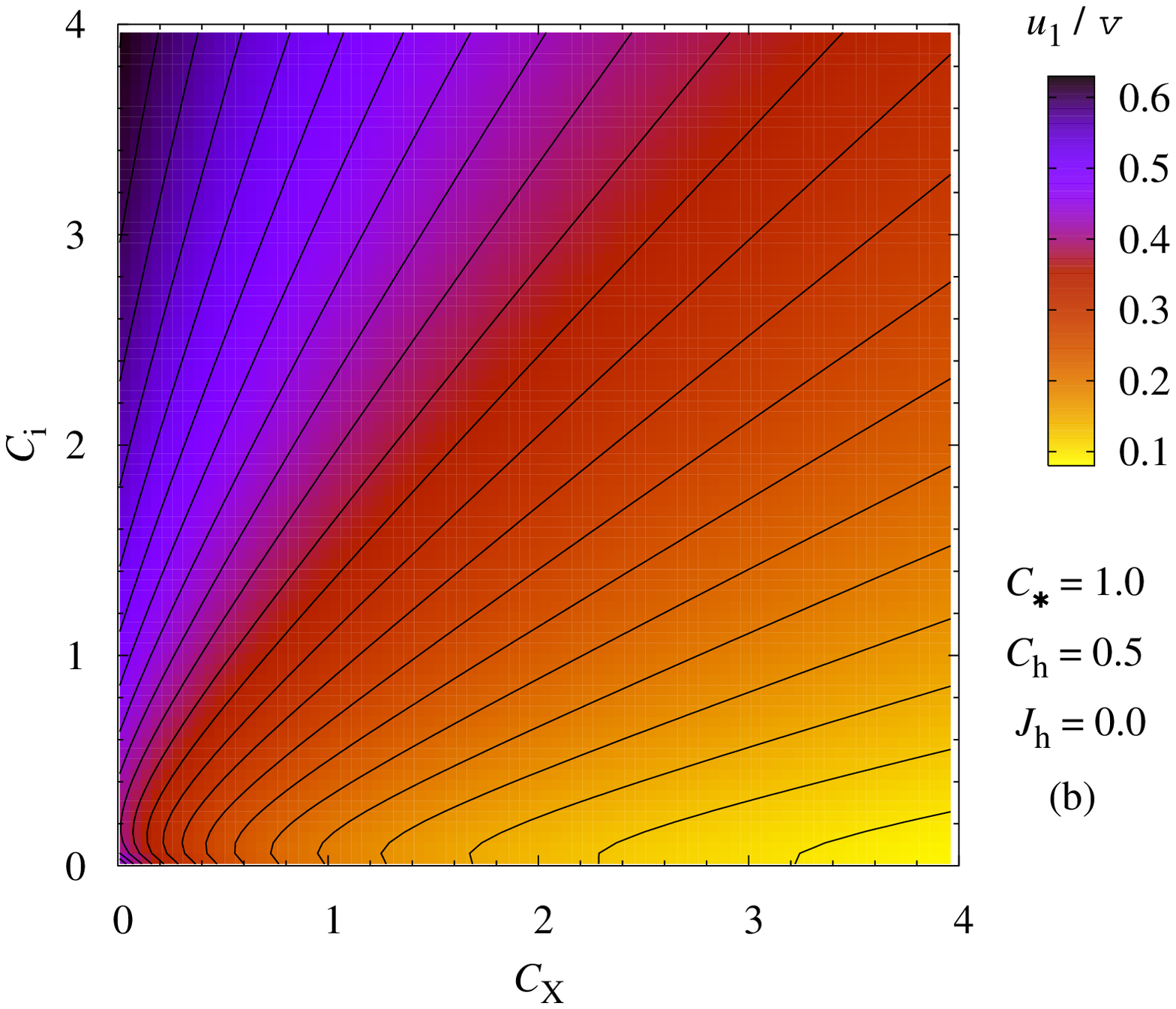}
\includegraphics*[width=\columnwidth]{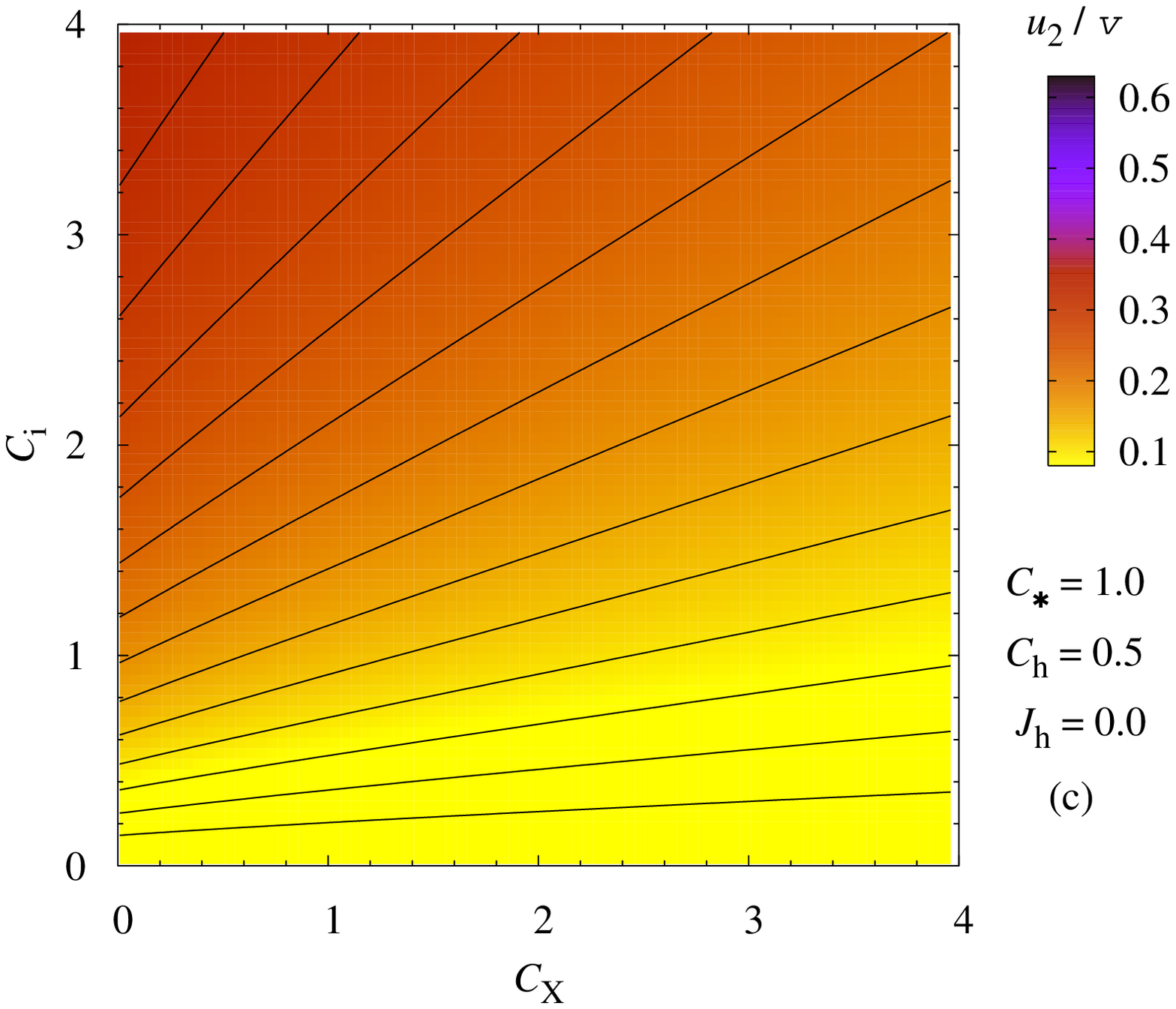}
\includegraphics*[width=\columnwidth]{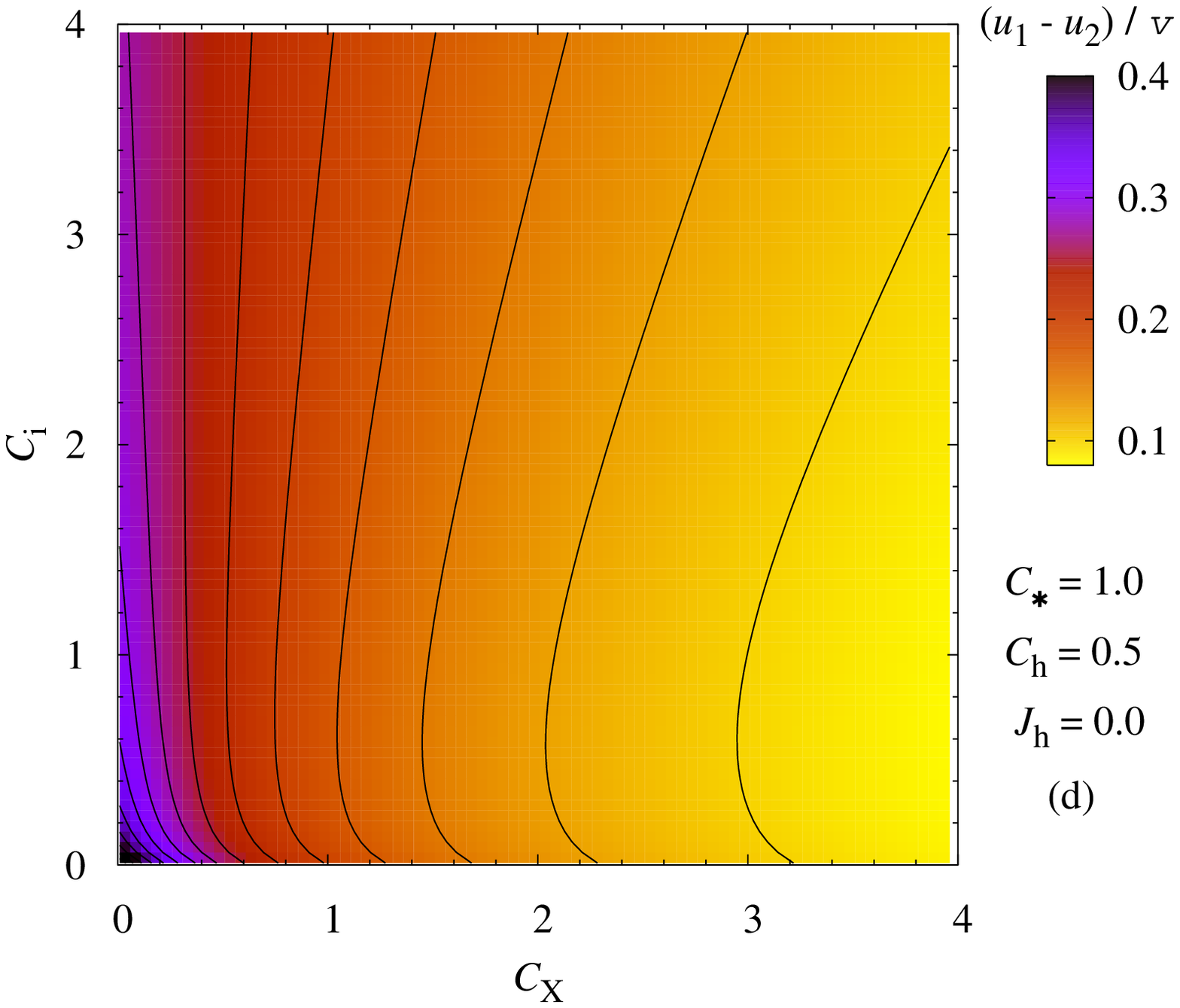}

\caption{(Color online)
Dependence of the amplitudes $v$, $u_1$ and $u_2$ on the capacitances
$C_{\rm X}$ and $C_{\rm i}$, for $C_{\ast} = 1$ and $C_{\rm h} =
0.5$. The results are obtained from the analysis of the $4 \times 4$
PBC cluster with $J_{\rm h} = 0$. Panels (b) and (c) suggest that
minimization of $u_1$ and $u_2$ is hard to be achieved and panel (d)
shows that our results for $v' = v - (u_1 - u_2)$ (our estimate for
$v$ for finite values of $J_{\rm h}$, see main text) underestimate $v$
by a factor of 40\% at most, and typically about 20\%.
}
\label{fig:u1u2}
\end{figure*}

Understanding the dependence of the interaction strengths on the
capacitances of the array is essential for  the optimization of the
energy scales in the emulated Hamiltonian, and for the minimization of
the couplings associated with the three-dimer interactions, the effect
of which are unknown.  Figure~\ref{fig:u1u2}  shows the dependence of
the couplings $v$, $u_1$, and $u_2$ on the capacitances \cite{Cstar}
on the same $4 \times 4$ cluster with PBC for $J_{\rm h} = 0$. The
interaction term $v(J_{\rm h} = 0)$ peaks for small values of $C_{\rm
i}$ and $C_{\rm X}$, and $u_1(J_{\rm h} = 0)$ and $u_2(J_{\rm h} =
0)$ are appreciable over a significant regime of the parameter space.
Decreasing $C_{\ast}$ leads to larger values of $u_1(J_{\rm h} = 0)$
and $u_2(J_{\rm h} = 0)$ (not shown) but has no significant effects
on $v(J_{\rm h} = 0)$.

To include the effects of nonzero Josephson current $J_{\rm h}$
on the interactions, we study the ten-hexagon cluster shown in
Fig.~\ref{fig:clusters}(b) using ENCORE. Due to the small size
of the cluster, only four-dimer configurations, out of a total of
14 in this cluster, give nonequivalent diagonal contributions in
the effective dimer Hamiltonian.  Boundary interactions due to open
boundary conditions further complicate the estimates.  Without going
to larger clusters (which is not possible with current computational
resources), we cannot determine the individual terms but only the
following combined quantity:
\begin{equation}
v' \equiv v - (u_1 - u_2)~.
\label{eq:v_prime}
\end{equation}
which is nevertheless a good estimate for $v$ as we can
see by analyzing the $4 \times 4$ cluster.  As shown in
Fig.~\ref{fig:u1u2}(d), $v'$ underestimates $v$ by typically only 20\%
for $J_{\rm h} = 0$ in the region where we have a valid mapping onto
the QDM, and these corrections do not change the conclusions drawn
below.\cite{current}

\subsection{Tuning the Ratio $v/t$}
\label{sec:ratio}

\begin{figure}[!tbp]
\centering
\includegraphics[width=0.35\textwidth,angle=270]{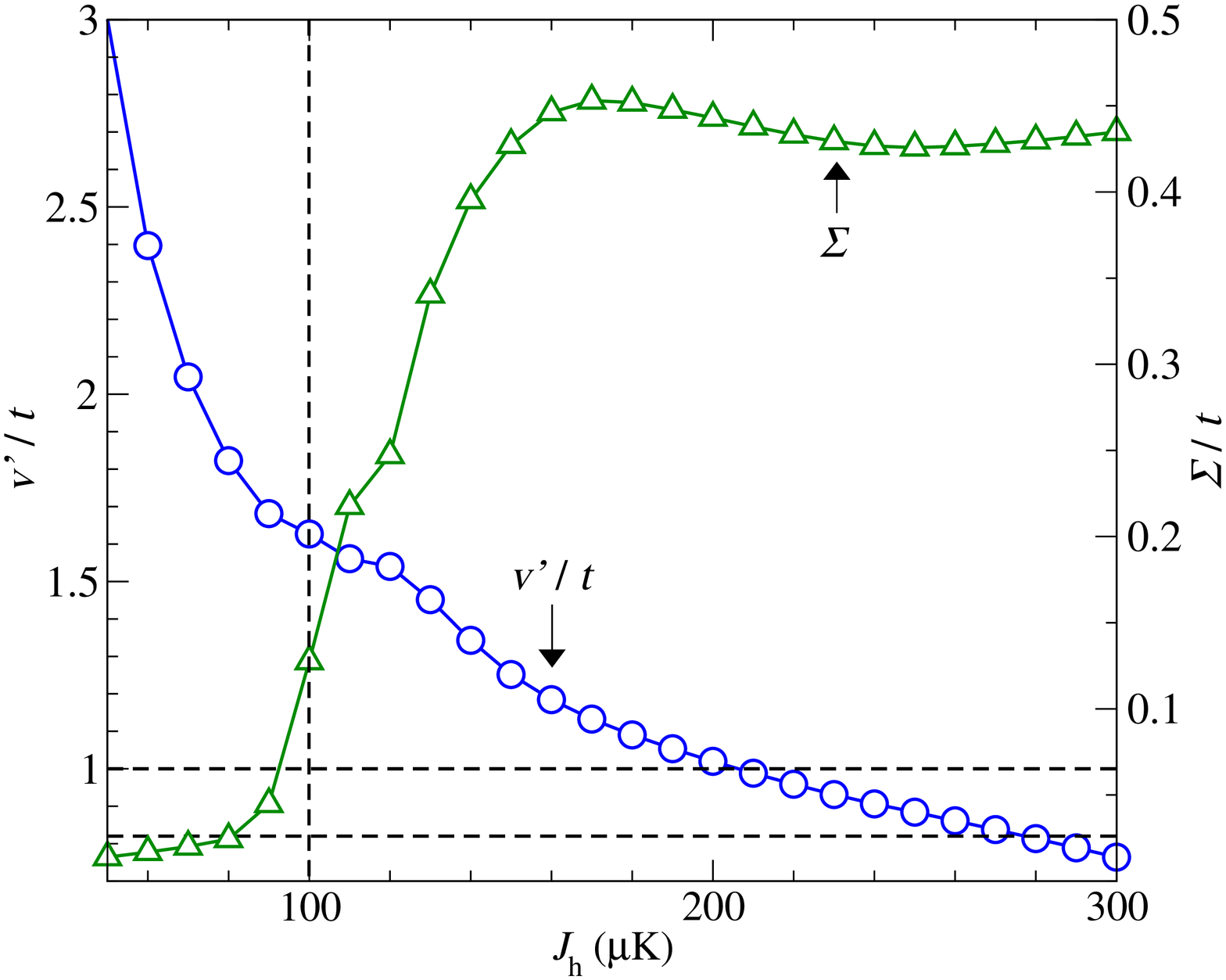}

\caption{(Color online) 
Our estimate $v' \equiv v - (u_1 - u_2)$ (circles)
(Ref.~\onlinecite{comparison}) for the two-dimer interaction $v$ vs
the Josephson current $J_{\rm h}$, for $C_{\ast} = C_{\rm X} = 10$,
$C_{\rm i} = 2$ and $C_{\rm h} = 0.5$, computed with ENCORE on the
ten-hexagon cluster. Triangles denote the sum of the absolute values
of the amplitudes ($\Sigma$) associated with {\em all} possible dimer
moves in the ten-hexagon cluster {\em beyond} the lowest order ones
depicted in Fig.~\ref{fig:clusters}(c)--\ref{fig:clusters}(e). The
breakdown of the mapping onto a dimer model, indicated by the vertical
dashed line, correlates with an abrupt increase in $\Sigma$ and occurs
before the topological phase of the simplest QDM for $0.82 \lesssim
v/t \le 1.0$ (indicated by the dashed horizontal lines) is reached
by $v'/t$.
}
\label{fig:vbyt01}
\end{figure}

Having discussed the dimer flips and interactions in the array of
Josephson junctions, we now analyze the feasibility of the emulation
of a topological phase by adjusting the couplings and required energy
scales in the emulated QDM to achieve an effective model with couplings
in the desired range $0.82 \lesssim v/t \le 1$.

In Fig.~\ref{fig:vbyt01} we show the ratio $v'/t$ as a function
of the Josephson current $J_{\rm h}$ for the set $C_{\ast}
= C_{\rm X} = 10$, $C_{\rm i} = 2$ and $C_{\rm h} = 0.5$ as
proposed in Ref.~\onlinecite{ioffe:02} (\onlinecite{comparison}). As seen
in Fig.~\ref{fig:vbyt01}, the topological phase corresponding to
$v'/t<1$  (marked by dashed horizontal lines)  is not reached before
the breakdown of the mapping to the QDM. This breakdown, marked by a
vertical dashed line is seen both  in the appearance of nondimer-like
states in the low-lying spectrum as well as by a drastic increase in
the summed amplitude for multi-dimer flips ($\Sigma$).  We conclude
that for this value of capacitances no topological phase exists.

\begin{figure*}[!tbp]
\includegraphics*[width=\columnwidth]{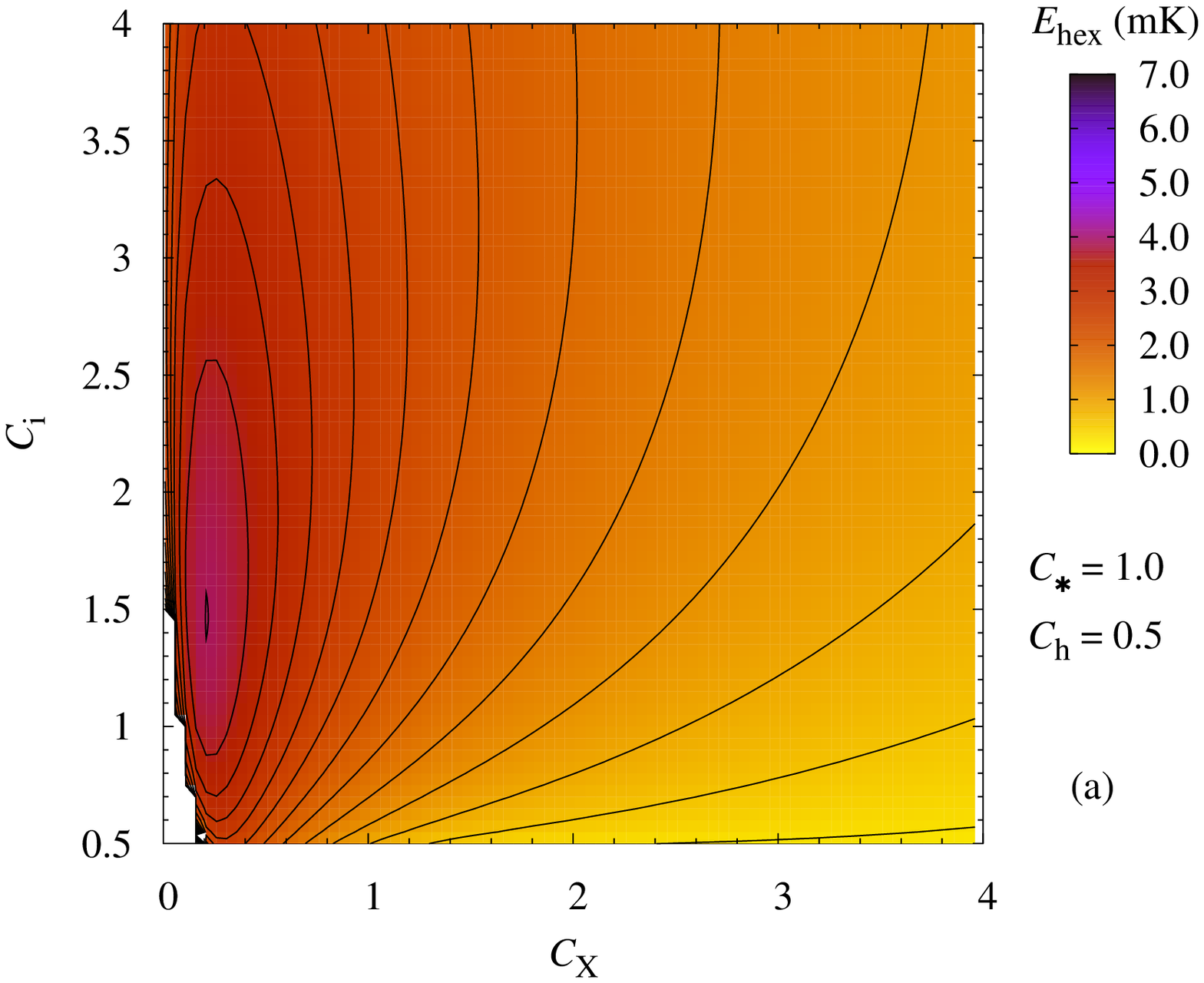}
\includegraphics*[width=\columnwidth]{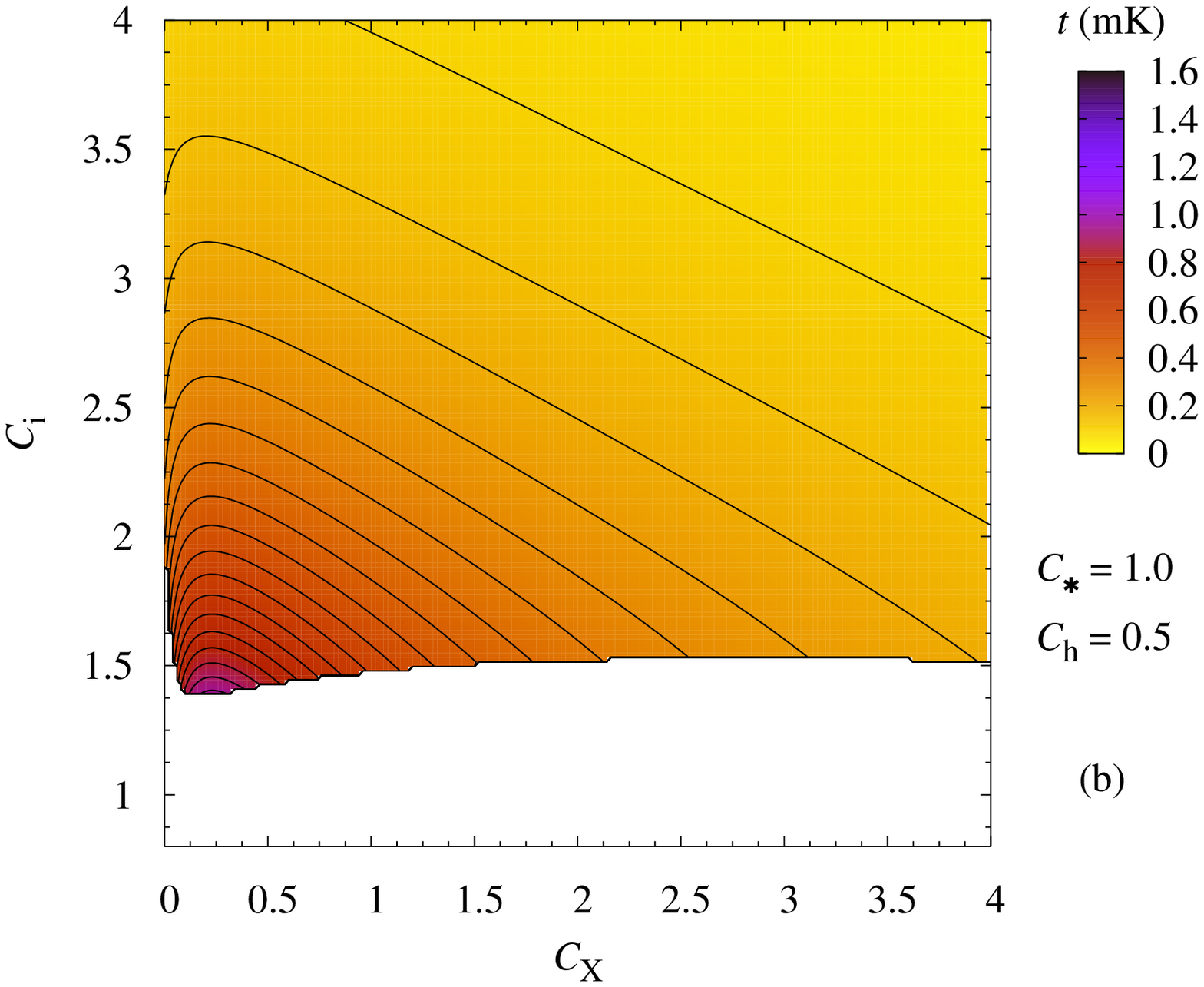}
\includegraphics*[width=\columnwidth]{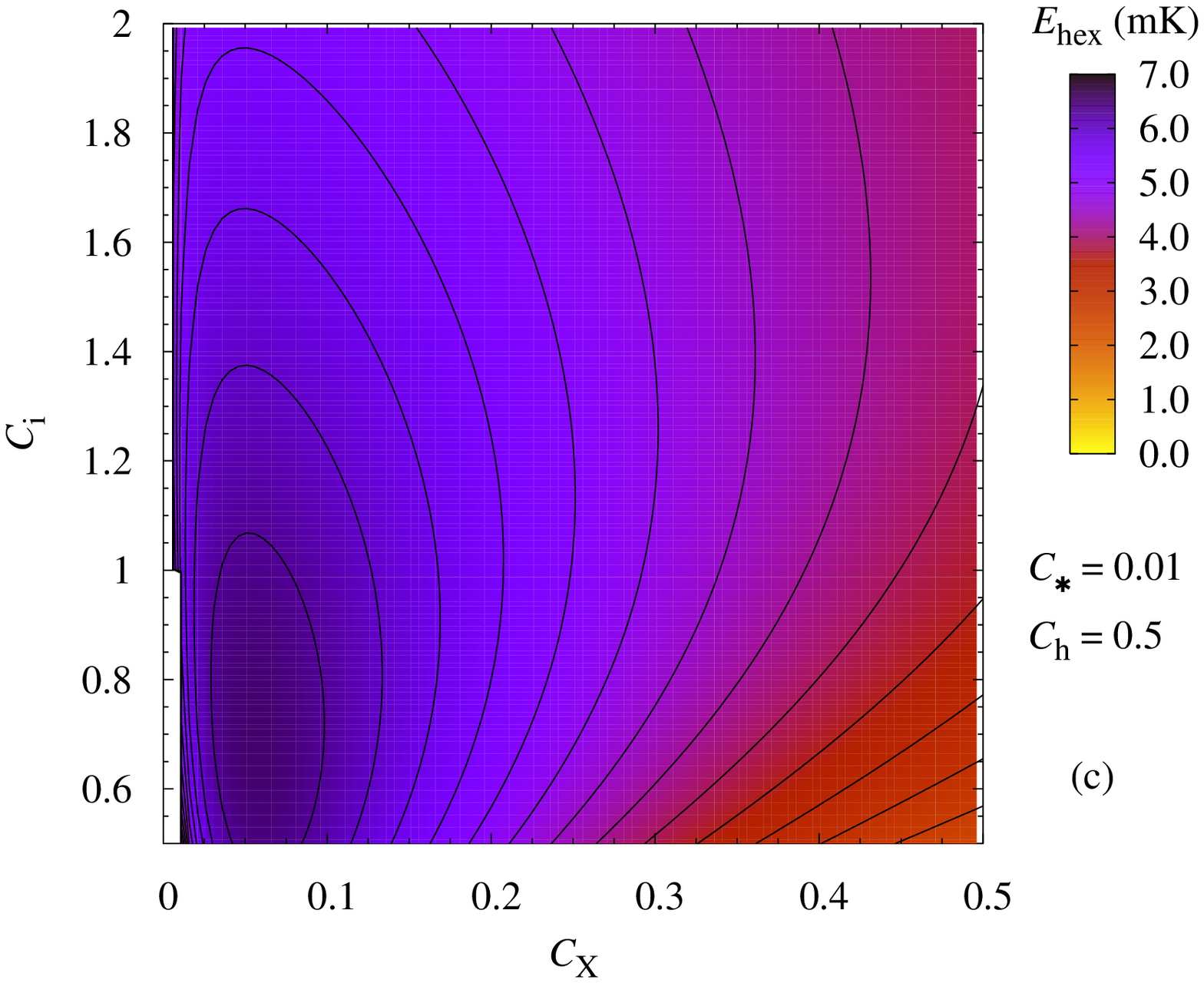}
\includegraphics*[width=\columnwidth]{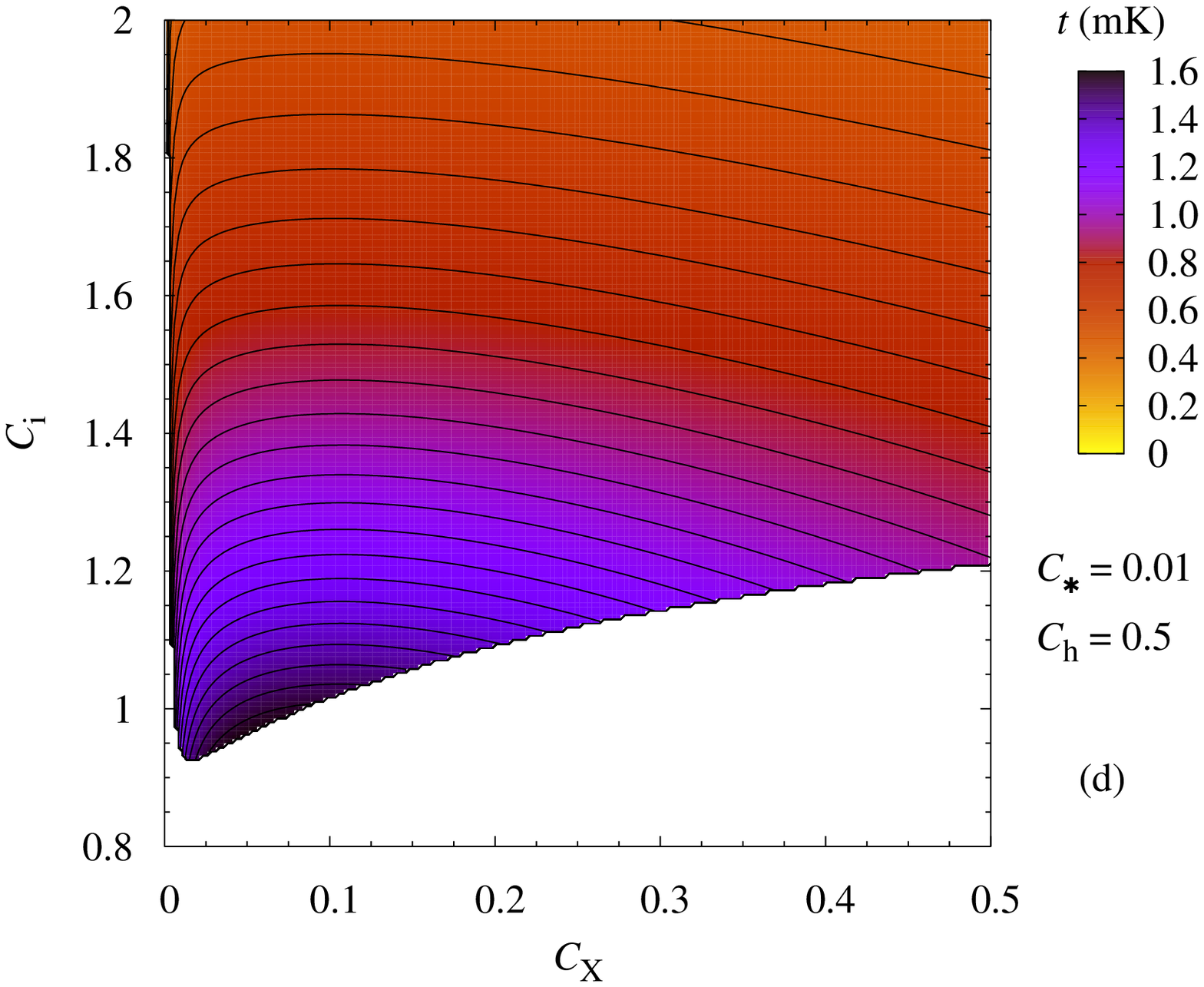}

\caption{(Color online)
The on-hexagon repulsion $E_{\rm hex}$ (left panel)
and the dimer flip amplitude  $t$ for $v/t = 1$, obtained by a second-order
perturbative analysis of the ten-hexagon cluster, as a function of the
capacitances $C_{\rm X}$ and $C_{\rm i}$ for $C_{\ast} = 1$ and $0.01$.
White regions in panels (a) and (c) indicate capacitances for which
$E_{\rm hex}$ is not a well defined quantity. In panels (b) and (d)
white regions correspond to sets of parameters for which the target
ratio for $v/t$ is reached for Josephson currents  $J_{\rm h} \geq
E_{\rm hex} / 2$, an upper bound for the largest current leading to a
valid mapping onto a dimer model.
}
\label{fig:Gap}
\end{figure*}

In order to stabilize the mapping onto a QDM for larger values of
the Josephson current, we explore alternative sets of capacitances
leading to larger values for $E_{\rm hex}$, more strongly suppressing
non-hardcore-dimer configurations.  The goal is to avoid the breakdown
of the mapping before the target values for $v/t$---where the system
is in the topological liquid phase---is reached.  The dependence
of $E_{\rm hex}$ on the capacitances $C_{\rm i}$ and $C_{\rm X}$
for two arbitrarily chosen values of $C_{\ast}$, obtained from a
numerical second-order perturbative analysis of the ten-hexagon
cluster  is shown in Figs.~\ref{fig:Gap}(a) and \ref{fig:Gap}(c).
There is a region close to the $C_{\rm i}$ axis where $E_{\rm hex}$
peaks, the peak value increasing  for smaller values of $C_{\ast}$.

Besides maximizing $E_{\rm hex}$, we also want to optimize the
amplitude $t$ since the topological phase appears only for temperatures
below the gap $\Delta \sim 0.1t$.  We use second-order numerical
perturbative results for the ten-hexagon cluster as a guide. For
each set of capacitances we calculate the value of the Josephson
current $J_{\rm h}$ giving the desired ratio $v/t = 1$, and we plot
the value of $t$ in Figs.~\ref{fig:Gap}(b) and \ref{fig:Gap}(d).
Since within perturbation theory, we are not able to determine whether
a particular set of parameters leads to a valid mapping onto a QDM,
we introduce an arbitrary cutoff, $J_{\rm h} = E_{\rm hex}/2$, which,
based on our ENCORE results is a generous upper bound. Capacitances for
which the target ratio $0.82 \lesssim v/t \le1$ is not reached below
this bound are discarded and indicated as blank regions in the figures.

These perturbative results show that optimal values for $t$ are
obtained in a region corresponding to small values of $C_{\rm X}$ and
close to the point where the breakdown of the mapping onto a QDM occurs
before the target ratio is reached. This behavior can be understood
if we analyze the dependence of $v(J_{\rm h} = 0)$ and $E_{\rm hex}$
on $C_{\rm X}$ and $C_{\rm i}$, shown in Figs.~\ref{fig:u1u2}(a),
Fig.~\ref{fig:Gap}(a), and \ref{fig:Gap}(c).  Since the ratio $v/t$ decreases
monotonically with $J_{\rm h}$, larger values of $v(J_{\rm h} =
0)$ have the desirable effect that values $0.82 \lesssim v/t \le 1$
are reached for larger values of $J_{\rm h}$, associated with more
favorable values for $t$ and $\Delta$.  Decreasing $C_{\rm i}$ at
small values of $C_{\rm X}$, we can see from Fig.~\ref{fig:u1u2}(a)
that progressively larger values of $v(J_{\rm h} = 0)$ can be obtained.
However, eventually, the region where $E_{\rm hex}$ peaks is surpassed
and the mapping onto a QDM breaks down before the target ratio
for $v/t$ is reached.  Decreasing $C_{\ast}$ does virtually not
affect $v(J_{\rm h} = 0)$ but considerably increases $E_{\rm hex}$,
therefore allowing us to obtain even more favorable values for $t$,
as shown in Fig.~\ref{fig:Gap}(d).

\begin{figure}[!tbp]
\includegraphics*[width=0.35\textwidth,angle=270]{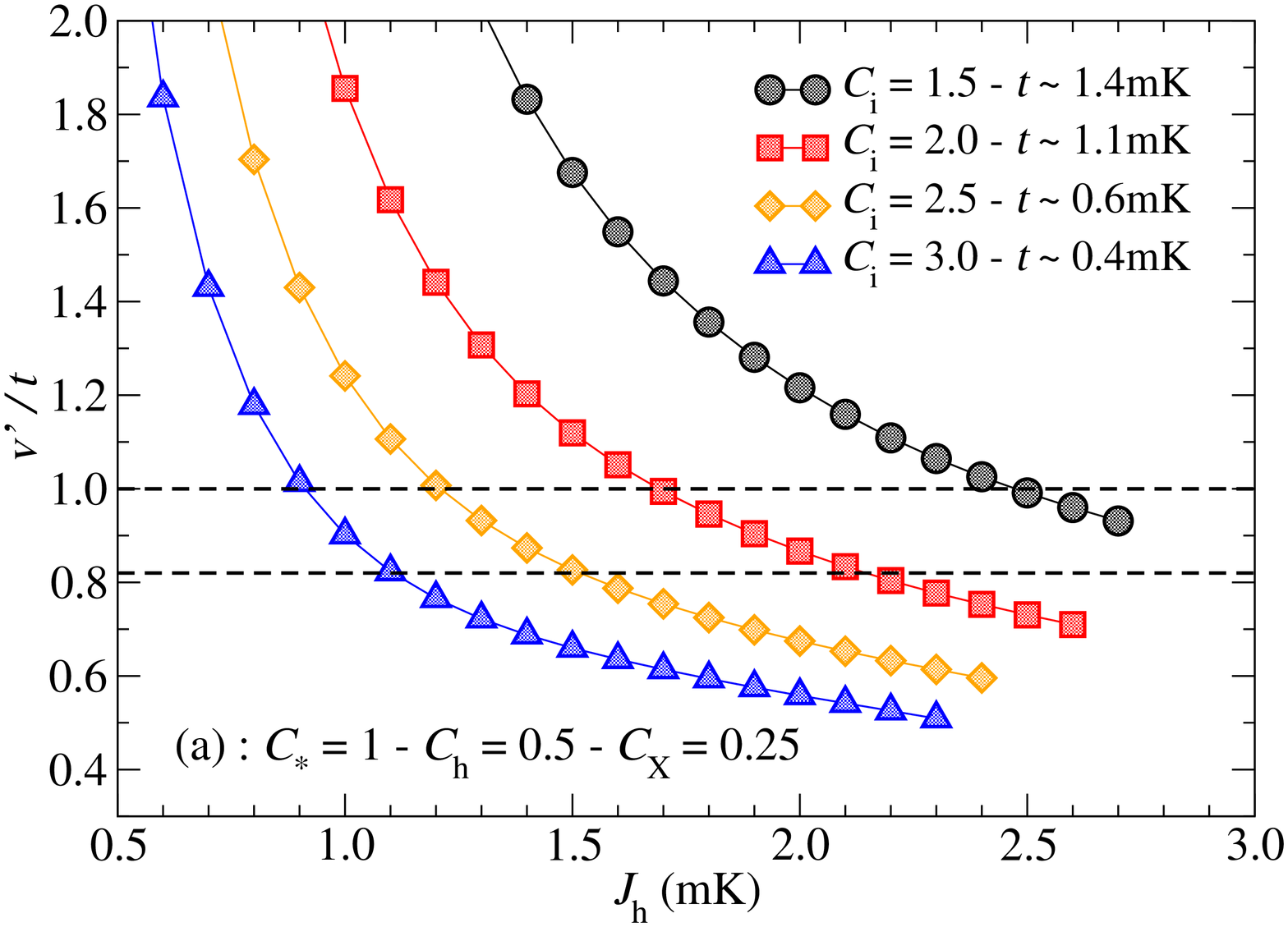}

\vspace{0.2cm}

\includegraphics*[width=0.35\textwidth,angle=270]{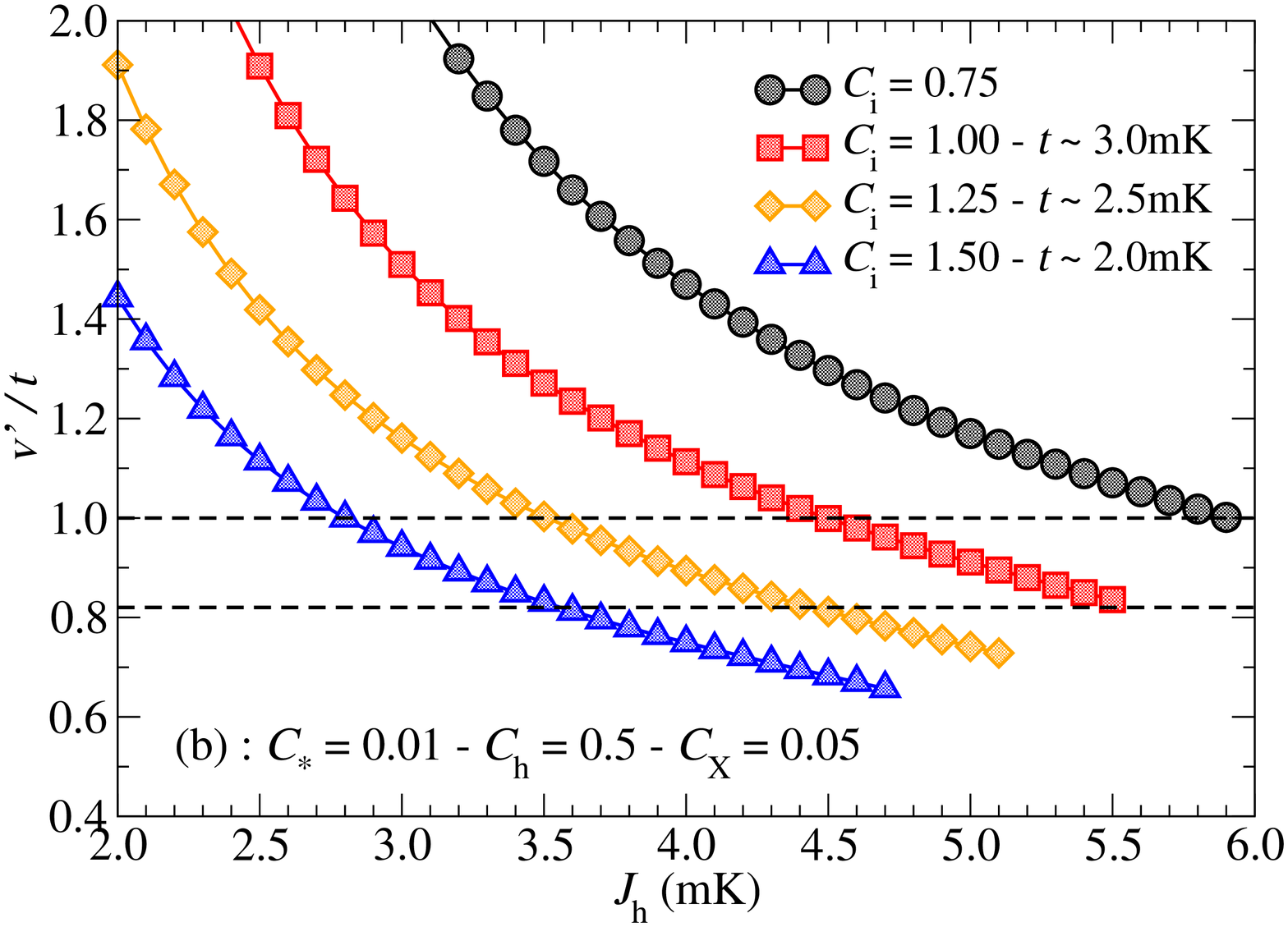}
\caption{(Color online) 
Dependence of the estimate for two-dimer repulsion $v'/t$ on the
Josephson current $J_{\rm h}$ and capacitance $C_{\rm i}$ for the
ten-hexagon cluster obtained with the ENCORE algorithm for (a)
$C_{\ast} = 1$, $C_{\rm X} = 0.25$, and $C_{\rm h} = 0.5$, and
(b) $C_{\ast} = 0.01$, $C_{\rm X} = 0.05$, and $C_{\rm h} = 0.5$
(Ref.~\onlinecite{zero_current}). The last data point on the right-hand
side corresponds to the value of $J_{\rm h}$ for which the mapping
onto a QDM breaks down. The dashed lines mark the spin-liquid phase
with TQO for the standard QDM.
}
\label{fig:vbyt_02}
\end{figure}

We have confirmed the validity of this qualitative analysis using
ENCORE.  In Fig.~\ref{fig:vbyt_02} we show  $v'/t$ as a function
of $J_{\rm h}$ for two different values of $C_{\ast}$. For each
value of $C_{\ast}$, $C_{\rm X}$ is chosen such that $E_{\rm hex}$
is close to its maximum and optimal values for $t$ can be obtained
(see Fig.~\ref{fig:Gap}). We terminate each curve when the mapping
breaks down, as seen by  nondimer intruder states in the low-energy
spectrum. In Fig. \ref{fig:vbyt_02}(a) we see that  for $C_{\ast} = 1$
and $C_{\rm X} = 0.25$ more favorable results for $t$ can be obtained
by decreasing  $C_{\rm i}$ , but below $C_{\rm i} \sim 1.5$ the mapping
onto a QDM breaks down before the target ratio $0.82 \lesssim v/t \le
1$ is reached. We thus get an upper bound of $t_{\rm max} \sim 1.5$mK
for the considered value of $C_{\ast}$. Larger values for the flip
amplitude can be obtained if we choose {\em smaller} $C_{\ast}$. But
as shown in Fig.~\ref{fig:vbyt_02}(b), only slightly larger values
of $t_{\rm max} \sim 3$mK are obtained for values of $C_{\ast}$ two
orders of magnitude smaller, suggesting that saturation is rapidly
reached.  Since experimentally we cannot arbitrarily decrease the
ground capacitances, we conclude that we can estimate an upper-bound
for the amplitude $t$ consistent with $0.82 \lesssim v/t \le 1$ of
only a few milli-Kelvin. These optimal values are associated with
small Josephson currents only slightly larger, $J_{\rm h} \lesssim
10$mK, typical experimental values being close to $1$K.  Returning to
the fact that $v'$ underestimates $v$ (Sec.~\ref{sec:interactions})
we see that this does not influence this upper bound estimate.

Thus, even if we assume that the longer-range terms present in the
emulated QDM do not destroy the topological liquid phase and that we
can still estimate the topological gap as $\Delta \sim 0.1t$, we can
expect that the operational temperatures for the putative quantum
bit is in the micro-Kelvin regime, clearly far below the limits of
current technologies.

\subsection{Extra Flips and Interactions}
\label{sec:extra}

\begin{figure}[!tbp]
\centering
\includegraphics*[width=0.35\textwidth,angle=270]{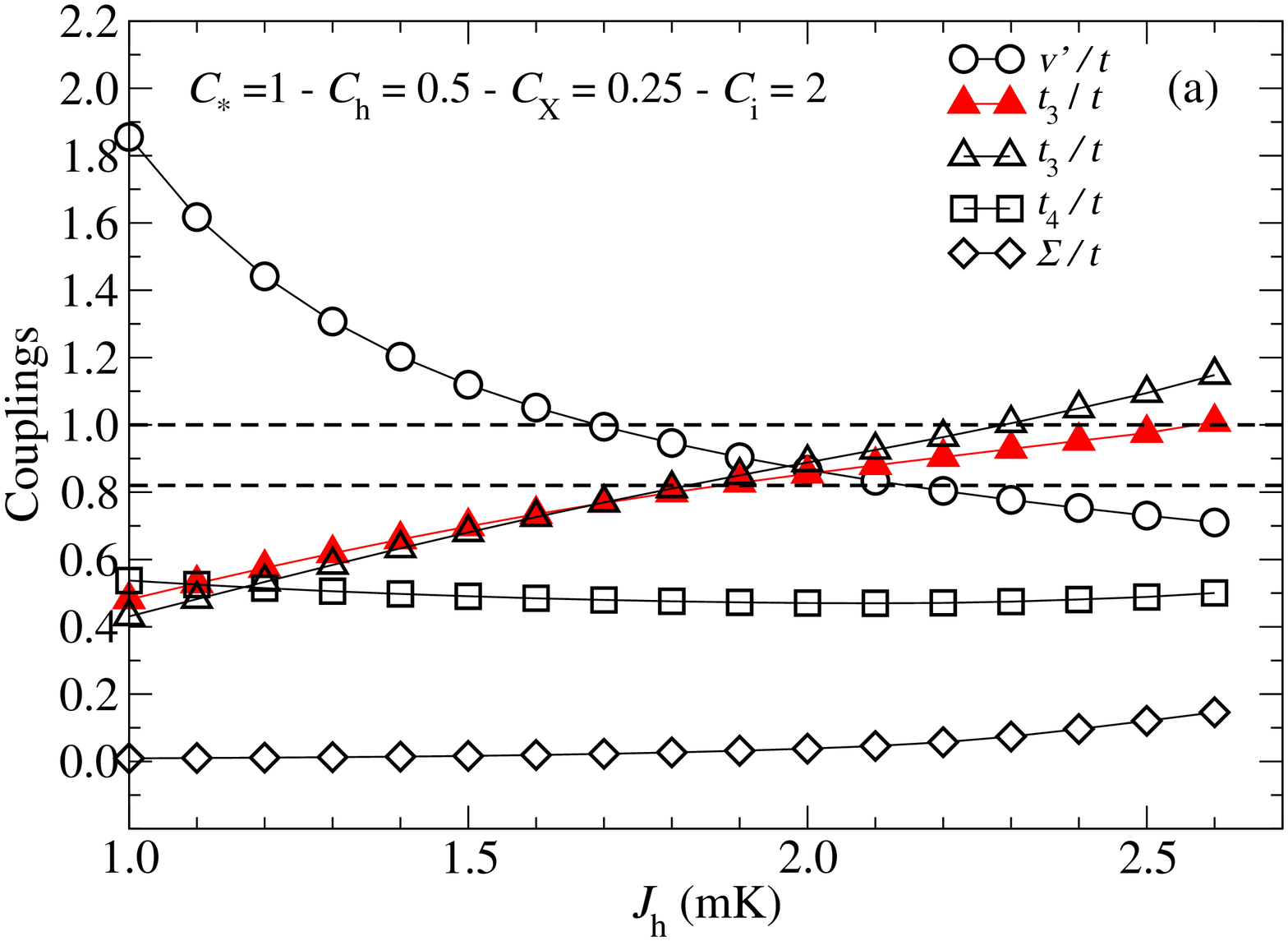}

\vspace*{0.2cm}

\includegraphics*[width=0.35\textwidth,angle=270]{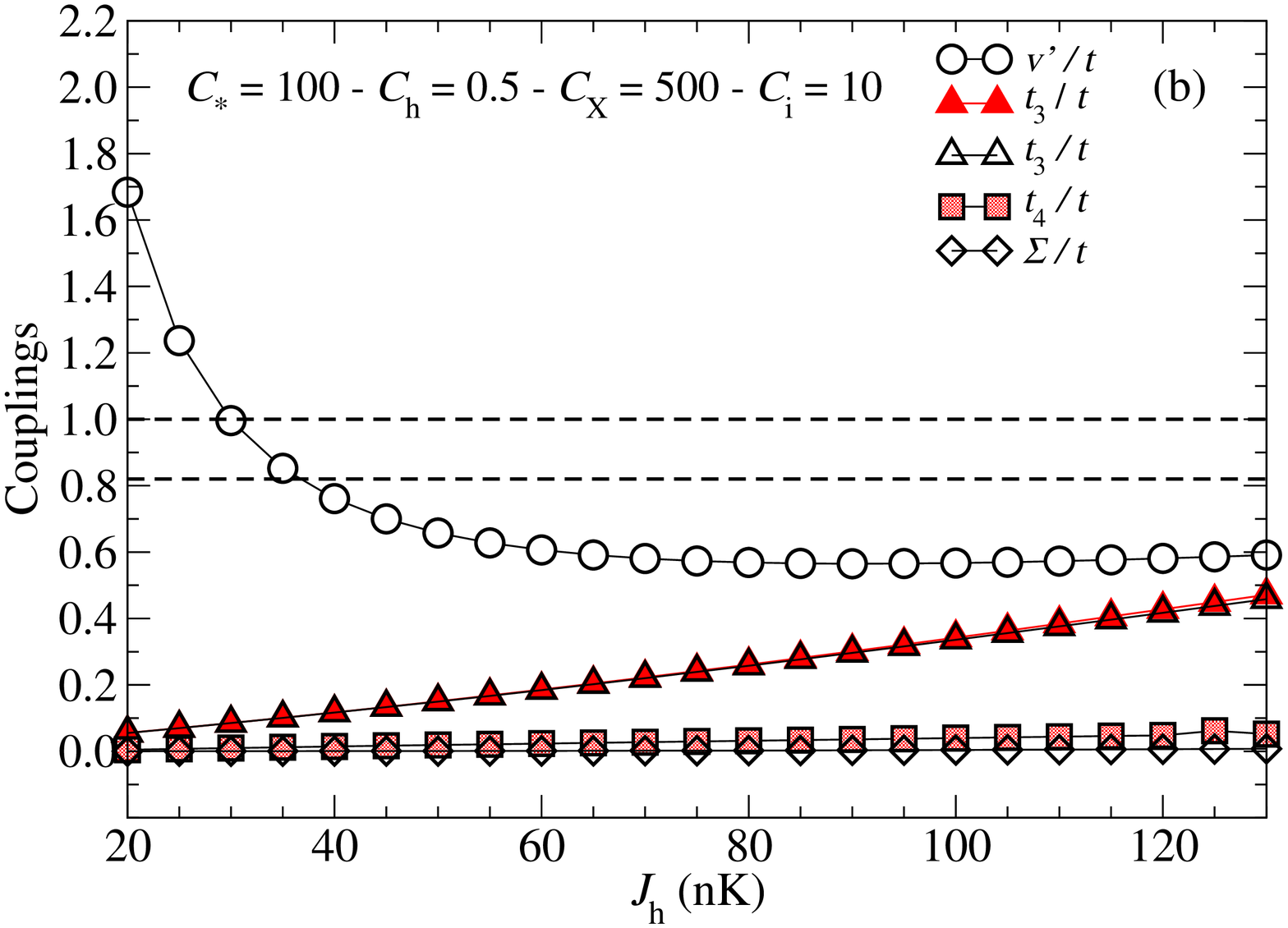}
\caption{(Color online)
Flip amplitudes and interactions obtained by ENCORE for (a) $C_{\ast} =
1$, $C_{\rm X} = 0.25$, $C_{\rm i} = 2$, and $C_{\rm h} = 0.5$ and (b)
$C_{\ast} = 100$, $C_{\rm X} = 500$, $C_{\rm i} = 10$ and $C_{\rm h}
= 0.5$.  In the upper panel only, the rightmost data point indicates
the breakdown of the mapping onto a QDM . Dashed horizontal lines
indicate the range of parameters corresponding to the topological
phase of the standard QDM.  Amplitudes for the special flips involving
three ($t_3$) and four ($t_4$) dimers are also shown. For $t_3$,
results are obtained from the analysis of the ten (filled triangles)
and six (empty triangles) hexagon clusters; $t_4$ is calculated from
the eight-hexagon cluster. In the limit $J_{\rm h}=0$ the strengths of
dimer interactions are: (a) $v=525\mu$K ($v'=390\mu$K), $u_1/v=0.512$,
and $u_2/v=0.253$ and (b) $v=0.879$nK ($v'=0.878$nK), $u_1/v=0.006$,
and $u_2/v=0.005$.  Values for $t$ corresponding to the target ratio
$0.82 \lesssim v/t \le 1.0$ are $t \sim1$mK for the set in panel (a)
and $t \sim 2$nK for the set in panel (b).
}
\label{fig:vbyt_03}
\end{figure}

So far we have ignored the presence of interactions and flips
comprising three or more dimers in the effective Hamiltonian, although
they are most likely relevant as suggested by the analysis of the
$J_{\rm h} = 0$ limit.\cite{zero_current} In Fig.~\ref{fig:vbyt_03}(a)
we show the dependence of the couplings on $J_{\rm h}$ for $C_{\ast}
= 1$, $C_{\rm X} = 0.25$,  $C_{\rm i} = 2$ and $C_{\rm h} = 0.5$.
In particular, the amplitudes associated with the unfrustrated flips
involving three and four dimers [Figs.~\ref{fig:clusters}(d) and
\ref{fig:clusters}(e)] are larger than the topological gap $\Delta
\sim 0.1t$ of the standard QDM. Vernay {\it et al.}\cite{vernay:06}
showed that the three-dimer flip extends the liquid phase.  Thus, in
order to be able to precisely estimate the operational temperatures
for the emulated qubit, it is necessary to study the effects of
the inclusion of the extra terms in the QDM.  However, even in the
absence of such a detailed analysis we can conclude that the involved
technological challenges in reaching the sub-milli-Kelvin temperatures
required for this device are substantial.

Aiming for a simpler QDM with only two-and three-dimer flips
[Figs.~\ref{fig:clusters}(c) and \ref{fig:clusters}(d)] requires the
suppression of higher-order flips. We find that this requires very
small values of $J_{\rm h}$ and leads to even smaller values of $t$.
To illustrate this problem, we show in Fig.~\ref{fig:vbyt_03}(b)
the dependence of all couplings on $J_{\rm h}$ in the emulated
QDM for $C_{\ast} = 100$, $C_{\rm X} = 500$, $C_{\rm i} = 10$, and
$C_{\rm h} = 0.5$.  With $J_{\rm h} = 0$ the interactions are given
by $v=0.879$nK, $u_1/v=0.006$, $u_2/v=0.005$, and $v'=0.878$nK;
nano-Kelvin temperatures are unrealistic in a solid-state device.

\section{Emulating Quantum Dimer Models Using Cold Atomic/Molecular Gases}
\label{sec:ca}

We now turn our attention to the alternative implementation based on
cold atomic/molecular gases loaded into an optical lattice which was
presented in Sec.~\ref{subsec:coldgas}. Since no concrete microscopic
proposal is available, we restrict ourselves to order of magnitude
estimates.

Flips comprising two dimers in this system similarly involve the
creation of a virtual state with energy $E_{\hexagon}$ and thus occur,
within second order in $J$, with amplitude $t = J^2 / E_{\hexagon}$
(we can also expect that flips involving three and four dimers
may play an important role here).  Since the mapping onto a QDM
necessarily breaks down when the kinetic energy dominates over the
on-hexagon repulsion, it follows that an upper bound for the hopping
amplitude consistent with a dimer picture is given by $J_{\rm max}
\approx E_{\hexagon} / 4$, and the largest obtainable value for the
flip amplitude in such emulator is thus $t_{\rm max} = J_{\rm max}^2/
E_{\hexagon} \approx J_{\rm max}/4$.

Preparation of the quantum bit state requires a controlled mixing
of dimer states belonging to different topological sectors. This
can be achieved  by virtually breaking one dimer and creating a
virtual particle-hole excitation, the particle corresponding to a
doubly-occupied hexagon and the hole to an empty one.\cite{ioffe:02}
For a qubit with linear dimension corresponding to $M$ hexagons,
an upper-bound for the mixing amplitude $h_x$ is given by
\begin{equation}
h_{x}^{\rm max} \sim 	J_{\rm max} 
			\left( \frac{J_{\rm max}}{2 E_{\hexagon}} \right)^{M-1} 
		= 	J_{\rm max} \left( \frac{1}{8} \right)^{M-1}.
\label{eq:hx}
\end{equation}

The largest attainable experimental values for the hopping amplitude in
cold atomic gases loaded into optical lattices are close to  $1$ kHz,
smaller values being expected for more massive molecules. Thus,
even on a rather small lattice comprised of the $10 \times 10$ hexagons
($M \approx10$), we can conclude that the time-scale involved in a
single qubit manipulation is of the order of minutes, much longer
than typical coherence times in cold atomic gases in optical lattices.

\section{Conclusions}
\label{sec:conclusions}

We have studied proposals to emulate a triangular lattice quantum
dimer model (QDM). A realistic emulation of the QDM would allow
the implementation of a fault-tolerant quantum bit, allowing us to
circumvent the problem of decoherence which plagues more conventional
proposals for achieving quantum computation.

The Josephson junction emulator by Ioffe {\em et
al.}\cite{ioffe:02}~was studied numerically using the ENCORE
method. Our results showed that the largest attainable values for the
two-dimer flip amplitude $t$ are a few milli-Kelvin and require very
small Josephson currents.  Since a device based on such an array would
only be operational at temperatures considerably below the topological
gap $\Delta \sim 0.1t$, implementation of a topologically-protected
quantum bit with the considered array is beyond the present day
technology.

The alternative array introduced by Ioffe and
collaborators\cite{ioffe:02} comprised of Y-shaped superconducting
islands forming a decorated triangular lattice would lead to even lower
values for the flip amplitudes since dimers in this implementation
correspond to a resonating Cooper pair and dimer flips involve
the tunneling of this pair through a weaker link with much smaller
Josephson current. Similar challenges with too low energy scales and
too long time scales are also faced by implementations using cold
atomic gases.

Our results illustrate the challenges involved in the design of
emulators for exotic phases. The fundamental reason for these
difficulties resides in the fact that topological quantum order
is a low-temperature feature, since the system's local degrees of
freedom must be highly entangled over long distances, of order of the
system's size, for topological order  to emerge. Since emulation of
the relevant models is obtained in the low-energy limit of the proposed
quantum device, we face the challenge that extremely low temperatures
are required. Thus, the approach of emulating topologically ordered
states for performing fault-tolerant quantum computation might only
be a successful one if we can devise emulators based on much stronger
bare electronic interactions, and a detailed analysis of engineering
limits is required.

\begin{acknowledgments}

A.F.A.~acknowledges discussions with I.~Milat and the financial support
from CNPq (Brazil) and NIDECO (Switzerland).  H.G.K.~acknowledges
support from the Swiss National Science Foundation under Grant
No.~PP002-114713.  We would like to thank A.~Wallraff for the fruitful
discussions. Part of the results shown in Fig.~\ref{fig:t_01} were
obtained by using the ALPS libraries (Refs.~\onlinecite{alet:05b}
and \onlinecite{albuquerque:07}).

\end{acknowledgments}

\bibliography{refs,comments}

\begin{thebibliography}{39}
\bibitem[{Cst()}]{Cstar}
\bibinfo{note}{We use now a value for $C_{\ast}$ smaller than the one used in
  Ref.~\onlinecite{ioffe:02}, since in this way it is possible to obtain larger
  values for $E_{\rm hex}$ and thus extend the mapping to larger values of
  $J_{\rm h}$, as discussed in Subsection \ref{sec:ratio}.}
\bibitem[{cur()}]{current}
\bibinfo{note}{Note that for finite $J_{\rm h}$ the quality of the estimate of
  $v$ improves.}
\bibitem[{com()}]{comparison}
\bibinfo{note}{For the set of capacitances $C_{\ast} = C_{\rm X} = 10$, $C_{\rm
  i} = 2.0$ and $C_{\rm h} = 0.5$, our estimate for the two-dimer interactions
  is $v'=29.5\mu$K when $J_{\rm h}=0$. These values should be compared with the
  ones obtained from the analysis of the $4\times4$ PBC cluster: $v=30.6\mu$K,
  $u_1/v=0.0842$ and $u_2/v=0.0478$.}
\end{thebibliography}

\end{document}